\documentclass[useAMS,usenatbib]{mn2e}
\usepackage{graphicx}
\usepackage[figuresright]{rotating}

\def\ngc{NGC\,6611}
\def\bd{BD\,-13$^{\circ}$}
\def\hd{HD\,168}

\def\tausco{$\tau$~Sco}

\def\l{~$\lambda$}
\def\ll{~$\lambda\lambda$}
\def\halp{H$\alpha$}

\def\halpha{H$\alpha$}

\def\hea{He\,{\sc i}}
\def\heb{He\,{\sc ii}}
\def\nc{N\,{\sc iii}}
\def\nd{N\,{\sc iv}}
\def\ne{N\,{\sc v}}
\def\mgb{Mg\,{\sc ii}}
\def\ob{O\,{\sc ii}}
\def\sic{Si\,{\sc iii}}
\def\sid{Si\,{\sc iv}}

\def\kms{km\,s$^{-1}$}
\def\cmss{cm\,s$^{-2}$}
\def\lsol{L$_{\odot}$}
\def\msol{M$_{\odot}$}
\def\rsol{R$_{\odot}$}
\def\s{$\sigma$}
\def\w{$\omega$}

\def\feros{{\sc feros}}
\def\flames{{\sc flames}}
\def\flgir{{\sc flames-giraffe}}
\def\gir{{\sc giraffe}}
\def\uves{{\sc uves}}
\def\fluves{{\sc flames-uves}}
\def\isis{{\sc isis}}
\def\hipparcos{{\sc hipparcos}}

\def\na{{\it n/a}}

\def\sss{\subsubsection}
\def\pdetect{$P_\mathrm{detect}$}

\defcitealias{SBL98}{SBL98}
\defcitealias{GM01}{GM01}
\defcitealias{HCB74}{HCB74}
\defcitealias{LM83}{LM83}
\defcitealias{PHYB90}{PHYB90}
\defcitealias{PHC91}{PHC91}
\defcitealias{BL95}{BL95}
\defcitealias{RCB97}{RCB97}
\defcitealias{BVF99}{BVF99}
\defcitealias{SGN08}{Paper~I}
\defcitealias{ESL05mnras}{ESL05}
\defcitealias{GPM07}{GPM07}

\def\hms{\citet{HMS93}}
\def\bmn{\citet{BMN99}}
\def\dse{\citet{DSE01}}
\def\esl{\citet{ESL05mnras}}

\def\phms{\citep{HMS93}}

\title[The massive star binary fraction in NGC\,6611]{The massive star binary fraction in young open clusters - II. NGC\,6611 (Eagle Nebula)}
\author[H. Sana et al.]{H. Sana$^{1,2}$\thanks{E-mail: hsana@eso.org}, 
   E. Gosset$^{3}$\thanks{FNRS, Belgium} and
   C. J. Evans$^4$\\
$^{1}$European Southern Observatory, Alonso de Cordova 1307, Casilla 19001, Santiago 19, Chile\\
$^{2}$Sterrenkundig Instituut Anton Pannekoek, Universiteit van Amsterdam, Postbus 94249, 1098 XH Amsterdam, The Netherlands\\
$^{3}$Astrophysical Institute, Li\`ege University, B\^at. B5c, All\'ee du 6 Ao\^ut 17, B-4000 Li\`ege, Belgium\\
$^{4}$UK Astronomy Technology Centre, Royal Observatory Edinburgh, Blackford Hill, Edinburgh EH9 3HJ, UK
}
\begin{document}

\date{Accepted 1988 December 15. Received 1988 December 14; in original form 1988 October 11}

\pagerange{\pageref{firstpage}--\pageref{lastpage}} \pubyear{2009}

\maketitle

\label{firstpage}

\begin{abstract}
Based on a set of over 100 medium- to high-resolution optical spectra collected from 2003 to 2009, we investigate the properties of the O-type star population in NGC\,6611 in the core of the Eagle Nebula (M16).  Using a much more extended data set than previously available, we revise the spectral classification and multiplicity status of the nine O-type stars in our sample. We confirm two suspected binaries and derive the first SB2 orbital  solutions for two systems. We further report that two other objects are displaying a composite spectrum, suggesting possible long-period binaries. Our analysis is supported by a set of Monte-Carlo simulations, allowing us to estimate the detection biases of our campaign and showing that the latter do not affect our conclusions. The absolute minimal binary fraction in our sample is $f_\mathrm{min}=0.44$ but could be as high as $0.67$ if all the binary candidates are confirmed. As in NGC\,6231 (see Paper I), up to 75\% of the O star population in NGC\,6611 are found in an O+OB system, thus implicitly excluding random pairing from a classical IMF as a process to describe the companion association in massive binaries. No statistical difference could be further identified in the binary fraction, mass-ratio and period distributions between NGC\,6231 and NGC\,6611, despite the difference in age and environment of the two clusters.
\end{abstract}


\section{Introduction}
At a distance of $\sim$2~kpc, NGC\,6611 at the core of the Eagle Nebula (M16) is one of the most famous star-formation regions of the Southern sky. Popularised by the dramatic HST image of \citet{HSS96mnras}, the striking morphology of the nebula is the result of the feedback of the massive stars in its core on the original natal cloud. In the last 6~Myr, M16  seems to have developed a continuous star-formation history, resulting in a diverse stellar population that encompasses high- and intermediate-mass stars, low-mass pre-main sequence stars, young embedded objects, maser sources and Herbig-Haro objects.  As such, the Eagle Nebula is also one of the best places to study the positive and negative influence of young massive stars to trigger new formation events. \citet{Oli09} has recently reviewed star formation in the Eagle Nebula, so that we only provide here some of the key facts that are of direct interest for the present work.

\begin{table}
\begin{minipage}{85mm}
\caption{Brief comparison of the properties of the various spectrographs used to assemble our data set. 
The last column provides the number of epochs $n$ obtained with each spectrograph.}
\label{tab: spectro}
\centering
\begin{tabular}{@{}lllc@{}}
\hline
Spectrograph & \ll\ range (\AA)    & Resolving power & $n$ \\
\hline
\feros      & 3800-9200            & 48\,000         & 78 \\
\flgir$^a$  & 3850-4755, 6380-6620 & 20\,850-29\,600 & 10 \\
\fluves     & 4140-6210            & 47\,000         & 6  \\
\uves       & 3280-4560            & 80\,000         & 3  \\
\isis       & 3800-5100, 6200-6800 & 7000            & 1  \\
\hline
\end{tabular}\\
\end{minipage}
$a.$ Each of the \gir\ epochs is actually composed of several observations using different wavelength settings. See \esl\ for more information.
\end{table}

\begin{table*}
 \begin{minipage}{170mm}
  \caption{Number of radial velocity (RV) measurements obtained for each object with respect to the considered spectral lines. `--' means that the corresponding line has not been measured. } \label{tab: sp_lines}
 \centering
  \begin{tabular}{@{}rrrrrrrrrrrrr@{}}
  \hline
         & \hea       & \sid       & \hea       & \hea       & \hea       & \mgb       & \heb       & \heb       & \hea     & \heb       & \hea       & \hea     \\
Object   & \l4026     & \l4089     & \l4144     & \l4388     & \l4471     & \l4481     & \l4542     & \l4686     & \l4922   & \l5412     & \l5876     & \l7065   \\
\hline                                                                                                                                
\bd4923   &  11       & 11         & 10         & 12         & 12         & 10         & 12         & 13         & 9        & 10         & 10         &  9       \\
\bd4927   &   7       &  7         & --         &  8         &  8         &  8         &  8         &  8         & 6        &  8         &  6         &  6       \\
\bd4928   &   7       &  7         &  7         &  8         &  8         & --         &  8         &  8         & 6        &  6         &  6         &  6       \\
\bd4929   &  11       & 11         & 11         & 11         & 12         & --         & 12         & 12         & --       & 10         & 10         & --       \\
\bd4930   &   7       &  7         &  7         &  8         &  8         &  8         &  8         &  8         &  6       &  6         &  6         &  5       \\
HD\,168075 &   9       &  9         &  9         &  9         & 11         & 11         & 11         & 11         & 11       & 11         &  9         &  8       \\
HD\,168076 &   9       & --         & --         & --         &  7         & --         &  7         &  7         & --       &  7         &  7         & --       \\
HD\,168137 &  14       & 14         & 14         & 14         & 14         & 14         & 14         & 14         & 13       & 14         & 14         & 12       \\
HD\,168183 &  13       & 13         & 13         & 13         & 13         & --         & 13         & 13         & 13       & 13         & 13         & 13       \\
\hline
\end{tabular}
\end{minipage}
\end{table*}

Several authors have identified variable extinction across the cluster and an abnormal reddening law. These properties have strongly affected the distance determination and a wide range of values, from 1.7 to 3.2~kpc, can be found in the literature. Yet, all the recent works seem to converge towards the closest distances. In the following, we adopt $d=1.8\pm0.1$~kpc as found by \citet{DSL06} from the spectroscopic parallaxes of 24 OB stars. 

The cluster age is 2-3 Myr typically, although the age spread is probably significant and extends from 1 to 6~Myr. The most recent age determinations \citep{BSB06,MFH08} seem to favour an even younger age around 1.5~Myr. Regarding the cluster mass, \citet{BSB06} argued that the OB-type stars alone amount for a minimal mass of $\sim1.6\times10^3$~\msol, although \citet{WSD07} suggested that the total mass could be as large as $\sim2.5\times10^4$~\msol.

As the second paper in our series, this work mainly focuses on the multiplicity properties of the O-type star population in NGC\,6611. Yet, given that those objects are the main source of radiative and kinetic energy of the Eagle Nebula, the present study has implications beyond the sole properties of the O-type stars in the cluster. It forms a prerequisite to any detailed understanding of the energetic balance of the mechanisms at work in the nebula.

 With 13 O-type stars and about 50 B0 to B5 stars, the young open cluster NGC\,6611 at the core of the Eagle Nebula hosts a rich early-type population. Compared to NGC\,6231 \citep[ \citetalias{SGN08}]{SGN08}, the O star population in NGC\,6611 is definitely younger and contains earlier spectral types.  While several authors have investigated the multiplicity of these objects, the picture is still not complete. \citet{BMN99} obtained multi-epoch low- ($R\approx4000$) and medium- to high-resolution ($R\approx9000-15000$) spectroscopy of ten of the brightest OB stars and reported three definite binaries and three multiple candidates. Subsequent medium- to high-resolution ($R\approx7000-29000$) spectroscopy by \citet{ESL05mnras} and \citet{MFH08} confirmed two of the known O-type binaries, further suggesting an additional SB1 candidate not observed by \citet{BMN99}. All in all, the minimum binary fraction $f_\mathrm{min}$ is 3/13$\approx$0.23. It could however be much higher given that some objects lack multi-epoch monitoring and  that four other binary candidates have been identified.

On the very large separation side, the adaptive optics survey of \citet{DSE01} shed a complementary light on the multiplicity properties of the cluster. Focusing in the separation range from 200 to 3000~A.U. ($\approx$0.1-1.5\arcsec) around 60 OB-type cluster members, they identified low-mass visual companions in 18$\pm$6\%\ of the cases, two of which are around O-type stars. 

Finally, \citet{GvB08} have recently identified three O-type stars that have most likely been ejected from NGC\,6611. Given the number of known O stars in the cluster (either isolated or within a multiple object), those ejected stars would amount to about 20\%\ of the initial O-type star population in NGC\,6611.  \citet{GvB08} further suggested that the three detected runaway stars might only represent about one fifth of the ejected O stars. This would imply that \ngc\ already displays a significant dynamical evolution. We emphasize that the present work only addresses the multiplicity properties of the current O-type star population in \ngc. 

 This work is organised as follows. Sect.~\ref{sect: obs} describes our observing campaign and the data reduction process. Sect.~\ref{sect: ostar} revises the properties of the individual O-type stars in the cluster. Sect.~\ref{sect: mc} presents an analysis of the observational biases using Monte-Carlo simulations. Finally, Sect.~\ref{sect: discuss} discusses the present results and Sect.~\ref{sect: ccl} summarizes our conclusions.


\setcounter{table}{+2}
\begin{sidewaystable*}
\begin{minipage}[t][180mm]{\textwidth}\begin{flushleft}
{\bf Table 3.} ~Journal of the spectroscopic observations of the O-type stars in NGC\,6611. First and second lines indicate the spectral line and the adopted rest wavelength (in \AA). The first column gives the heliocentric Julian date at mid-exposure. The following columns provide, for each spectral line, the measured RVs (in \kms).  References for the instrumental setup can be found at the bottom of the table. The full table is available in the electronic edition of the journal.\end{flushleft}
 \centering
\begin{tabular}{@{}rrrrrrrrrrrrr@{}}
\hline												
HJD           & \hea\l4026 & \sid\l4089 & \hea\l4144 &\hea\l4388 & \hea\l4471 & \mgb\l4481 & \heb\l4542 & \heb\l4686 & \hea\l4922 &\heb\l5412 &\hea\l5876 & \hea\l7065 \\
$-2\,400\,000$& 4026.072   &  4088.863  &  4143.759  & 4387.928  &  4471.512  &  4481.228  &  4541.590  &  4685.682  &  4921.929  & 5411.520  & 5875.620  & 7065.190   \\
\hline 
\multicolumn{13}{c}{\bd4923 prim}\\
\hline
52836.70812$^{d}$     &  --     &  --     &    82.4 &  --     &    62.7 &    55.1 &    68.6 &    91.7 &  --     &  --     &  --     &  --  \\
52839.75137$^{d}$     &  --     &  --     &  --     &  --     &    60.5 &    52.6 &    75.2 &    92.2 &  --     &  --     &  --     &  --  \\
53134.95474$^{e}$     &  --     &  --     &  --     &  --     &  --     &  --     &  --     &    11.2 &  --     &    11.3 &    42.0 &  --  \\
53509.84538$^{a}$     & $-$79.4 &  --     &  --     &  --     & $-$63.7 &  --     & $-$64.4 & $-$56.3 &  --     & $-$63.8 & $-$70.5 &  --  \\
53510.87032$^{a}$     &  --     &  --     &  --     &  --     &  --     &  --     &  $-$4.1 & $-$11.5 &  --     &  $-$0.2 &    24.6 &  --  \\
53511.76887$^{a}$     &  --     &  --     &  --     &  --     &  --     &  --     &    18.2 &    25.0 &  --     &    18.4 &  $-$2.9 &  --  \\
53512.75277$^{a}$     &  --     &  --     &  --     &  --     &    47.3 &    44.6 &    35.6 &    49.9 &  --     &    37.9 &    54.9 &  --  \\
53860.83796$^{a}$     &    80.7 &  --     &  --     &  --     &    58.8 &  --     &    69.0 &    81.9 &  --     &    70.3 &    62.2 &  --  \\
53861.72041$^{a}$     &  --     &  --     &    31.8 &  --     &    43.4 &    56.1 &    39.1 &    78.7 &    49.7 &    41.3 &    56.2 &  --  \\
53862.72524$^{a}$     &  --     &  --     &    22.0 &  --     &  --     &    46.2 &    27.5 &    59.5 &  --     &    27.2 &    42.7 &  --  \\
53863.84924$^{a}$     &  --     &  --     &  --     &  --     &  --     &  --     &    16.3 &    42.2 &  --     &    14.2 &     6.3 &  --  \\
53864.75611$^{a}$     &  --     &  --     &  --     &  --     &  --     & $-$43.6 &  $-$3.5 &    24.9 & $-$52.7 &  $-$2.1 &    17.8 &  --  \\
\hline
\multicolumn{13}{c}{\bd4923 sec}\\
\hline
52836.70812$^{d}$     & $-$21.1 & $-$85.5 & $-$83.0 & $-$79.8 & $-$86.0 & $-$88.3 &$-$126.1 & $-$80.5 &  --     &  --     &  --     &  --     \\
52839.75137$^{d}$     &  --     &  --     &  --     & $-$81.5 & $-$86.3 & $-$89.1 &$-$101.5 & $-$76.6 &  --     &  --     &  --     &  --     \\
53134.95474$^{e}$     &  --     &  --     &  --     &  --     &  --     &  --     &  --     &  --     &  --     &  --     & $-$50.6 &  --     \\
53509.84538$^{a}$     &   161.0 &   157.8 &   149.1 &    56.0 &   152.3 &   142.0 &   164.0 &   158.3 &   149.8 &   174.8 &   153.9 &  53.2   \\
53510.87032$^{a}$     &     4.3 &    63.5 &    64.2 &    60.9 &    15.1 &  --     &  --     &    49.9 &    55.2 &  --     &  --     &  $-$9.6 \\
53511.76887$^{a}$     &  $-$9.1 & $-$21.5 & $-$27.6 & $-$23.1 & $-$12.2 & $-$14.3 &  --     & $-$11.5 & $-$19.6 &  --     &  --     &  $-$9.6 \\
53512.75277$^{a}$     & $-$27.0 & $-$69.3 & $-$73.9 & $-$68.5 & $-$74.2 & $-$81.1 &  --     & $-$82.5 & $-$70.8 &  --     & $-$71.7 & $-$71.5 \\
53860.83796$^{a}$     & $-$85.0 & $-$93.0 &  --     & $-$97.7 &$-$101.5 &$-$102.1 &$-$139.1 & $-$95.8 & $-$96.7 &$-$115.1 & $-$96.1 & $-$99.3 \\
53861.72041$^{a}$     & $-$19.9 & $-$79.2 & $-$80.9 & $-$74.3 & $-$88.9 & $-$84.4 &  --     & $-$88.3 & $-$81.1 &  --     & $-$77.1 & $-$79.5 \\
53862.72524$^{a}$     & $-$13.5 & $-$49.1 & $-$50.3 & $-$46.0 & $-$32.8 & $-$50.9 &  --     & $-$67.2 & $-$44.4 &  --     & $-$46.4 & $-$44.4 \\
53863.84924$^{a}$     &  $-$6.7 &     3.6 &     5.5 &  $-$2.6 &     2.9 &  $-$1.7 &  --     & $-$32.6 &     3.4 &  --     &  --     &     6.6 \\
53864.75611$^{a}$     &  $-$0.1 &    62.5 &    56.3 &    58.0 &    16.4 &    46.7 &  --     &  --     &    57.9 &  --     &  --     &    54.1 \\
\hline
\end{tabular}\\
\begin{flushleft}$a.$ ESO2.2m + FEROS ; $b.$ VLT + UVES; $c.$ VLT + FLAMES-UVES; $d.$ VLT + FLAMES-GIRAFFE; $e.$ WHT + ISIS 
\end{flushleft}
\end{minipage}
\end{sidewaystable*}


\setcounter{table}{+3}
\begin{table*}
 \centering
 \begin{minipage}{160mm}
  \caption{Diagnostic line ratios and corresponding spectral types for the O-type objects studied. \na\ indicates criteria that either do not apply to the considered spectral type or for which the lines could not be securely measured and/or separated. We refer to text for discussion of the individual cases and to Table~\ref{tab: bin} for the finally adopted spectral types. The error-bars give the 1-\s\ dispersions on the measurements.} 
\label{tab: EW}
  \begin{tabular}{@{}llrrrrl@{}}
  \hline
Object     & Component     & $\log W'$        & $\log W''$      &$\log W_{\lambda4686}$& $\log W'''$ &  Sp. Type\\
  \hline     
\bd4923    & Primary       & $-0.612\pm0.130$ &  \na             & $>2.751\pm0.071$ & \na             & O4~V((f))  \\
\bd4923    & Secondary     & $ 0.125\pm0.063$ &  \na             & $>2.147\pm0.182$ & \na             & O7.5~V     \\
\bd4927    & \na           & $-0.020\pm0.014$ &  \na             & $ 2.348\pm0.041$ & \na             & O7~III(f)  \\
\bd4928    & \na           & $ 0.541\pm0.057$ &  $0.147\pm0.151$ & \na              & $5.531\pm0.109$ & O9.5~V     \\
\bd4929    & Primary       & $-0.044\pm0.086$ &  \na             & $>2.637\pm0.063$ & \na             & O7~V       \\
\bd4930    & \na           & $ 0.272\pm0.066$ & $-0.299\pm0.021$ & \na              & $5.333\pm0.009$ & O8.5~V     \\
HD\,168075  & Primary       & $-0.096\pm0.030$ &  \na             & $>2.837\pm0.012$ & \na             & O6.5~V((f))\\
HD\,168076  & Composite     & $-0.721\pm0.061$ &  \na             & \na              & \na             & O4~V((f))  \\
HD\,168137  & Primary       & $-0.075\pm0.039$ &  \na             & $>2.644\pm0.033$ & \na             & O6.5~V     \\
HD\,168137  & Secondary     & $ 0.202\pm0.068$ &  \na             & $>2.598\pm0.046$ & \na             & O8~V       \\
HD\,168183  & Primary       & $ 0.608\pm0.026$ &  $0.034\pm0.031$ & \na              & $>5.270\pm0.021$ & O9.5~III   \\
\hline
\end{tabular}
\end{minipage}
\end{table*}

\begin{table*}
 \centering
 \begin{minipage}{170mm}
  \caption{Cross-identification and physical properties of the O-type stars in \ngc. The magnitudes are from \citet{GPM07}, except for the two brightest objects for which data  from \citet{HMS93} are quoted. The stellar parameters come from \citet{DSL06}. The uncertainties are the 1-\s\ error-bars quoted by the authors.}
\label{tab: ID}
  \begin{tabular}{@{}lllllllllllll@{}}
  \hline
HD     & BD   & \citeauthor{Wal61}   & \citetalias{ESL05mnras} & \citetalias{GPM07} &   $V$            & $B-V$           & $T_\mathrm{eff}$ & $\log g$  & $\log L$ &$v \sin i$  \\  
       &      &                      &                         &                    &                  &                 &    (kK)         & (\cmss)   & (\lsol)  & (\kms)     \\  
  \hline                                                                                                                                                                                                                 
168183$^a$ & $-$13$^{\circ}$4991 & W412 &  6611-001               & 24374              &  8.18            & 0.34            & 32.0             & 3.60      & 5.39     & 142        \\  
168076$^a$ & $-$13$^{\circ}$4926 & W205 &  6611-002               & 27436              &  8.18            & 0.43            & 41.5             & 3.90      & 5.86     & 102        \\  
168075$^a$ & $-$13$^{\circ}$4925 & W197 &  6611-003               & 18360              &  8.752$\pm$0.037 & 0.543$\pm$0.107 & 40.0             & 3.90      & 5.63     &  87        \\  
168137$^a$ & $-$13$^{\circ}$4932 & W401 &  6611-004               & 13706              &  8.942$\pm$0.006 & 0.407$\pm$0.025 & 37.0             & 4.00      & 4.85     &  76        \\  
    -- & $-$13$^{\circ}$4930$^b$ & W367 &  6611-006               & 13867              &  9.368$\pm$0.023 & 0.343$\pm$0.079 & 34.0             & 4.10      & 4.84     &  43        \\  
       &                      &      &                         &                    &                  &                 & 31.3             & 4.00      & 4.81     &  20        \\  
    -- & $-$13$^{\circ}$4927     & W246 &  6611-008               & 6835               & 11.114$\pm$0.541$^c$ & 0.669$\pm$0.607 & 36.0             & 3.50      & 5.75     & 100        \\  
    -- & $-$13$^{\circ}$4929$^a$ & W314 &  6611-011               & 19208              &  9.803$\pm$0.203 & 0.706$\pm$0.208 & 36.0             & 4.20      & 5.24     &  66        \\  
    -- & $-$13$^{\circ}$4923     & W175 &  6611-014               & 5890               & 10.007$\pm$0.011 & 0.979$\pm$0.019 &  --              &   --      &   --     &  --        \\  
    -- & $-$13$^{\circ}$4928     & W280 &  6611-015               & 1806               & 10.044$\pm$0.039 & 0.560$\pm$0.048 & 32.5             & 3.90      & 4.77     & 410        \\  
    -- &               --       & W166 &  6611-017               & 18715              & 10.296$\pm$0.003 & 0.728$\pm$0.010 & 36.0             & 3.95      & 5.02     &  95        \\  
    -- &               --       & W161 &  6611-029               & 5818               & 11.215$\pm$0.007 & 1.241$\pm$0.008 & 36.0             & 3.85      & 5.37     & 135        \\  
    -- &               --       & W584 &  6611-045               & 5510               & 12.052$\pm$0.008 & 1.235$\pm$0.010 & 35.0             & 4.00      & 5.04     &  25        \\  
    -- &               --       & W222 &  6611-080               & 3820               & 12.968$\pm$0.008 & 1.519$\pm$0.009 & 40.0             & 4.00      & 5.33     &  95        \\  
\hline              
                                
\end{tabular}\\
$a.$ For those objects, the composite nature of their spectrum was not known, thus not taken into account while deriving the stellar parameters. The latter might thus be indicative only.\\
$b.$ The entries on the second row quote the results obtained by \citet{HDS07}, who re-analysed the \bd4930 spectrum and obtained slightly different stellar parameters.\\
$c.$ \citet{HMS93} quote $V=9.46$, suggesting that BD$-$13$^{\circ}$4923 is brighter than reported by \citet{GPM07}, although marginally in agreement within the (rather large) error-bars from the latter.\\
\end{minipage}
\end{table*}

\section{Observations and data handling} \label{sect: obs}

The core of our data set is formed by 66 high-resolution  \feros\ spectra obtained from May 2004 to June 2006 at the ESO/MPG-2.2m telescope at La Silla (PI: Sana). The \feros\ spectrograph provides us with the complete optical spectrum (3800-9200~\AA) at once, with a spectral resolving power of 48\,000. The \feros\ properties are identical to those given in \citetalias{SGN08}, and the data reduction process is described in detail in \citet{San09}. Exposure times ranged between 10 and 45~min depending on the object magnitude, yielding a typical signal-to-noise ratio (SNR) of 200 as measured in the continuum close to 5000~\AA. 

A second part of our data set is composed of 10 \flames-\gir\ spectra, five \feros\ spectra and six \fluves\ spectra from the VLT-\flames\ survey of massive stars (PI: Smartt). Those data and the data reduction have been previously described by \citet{ESL05mnras}.  Additionally, we retrieved  one \feros\ spectrum of HD\,168075 and one of HD\,168076 (PI: Bouret) from the ESO archive. We also retrieved three UVES long-slit echelle spectra of HD\,168076 (PI: Andre). The two \feros\ spectra were reduced as described above while the \uves\ data were reduced using the ESO CPL-based pipeline and normalised by fitting low-order polynomials to the continuum of the individual orders.

Finally, five objects were re-observed with \feros\ in March 2009 (PI: Evans) to improve the detection likelihood of long-period systems. These additional data were reduced as described above and have a typical SNR of 250 to 300 at 5000~\AA. Table~\ref{tab: spectro} gives a brief overview of the wavelength coverage and spectral resolution of the different spectra in our data set. 

Using the reduced spectra, Doppler shifts and equivalent widths (EWs) were measured by simultaneously fitting one to three Gaussians (depending on whether an SB1, SB2 or SB3 signature was visible) to a series of line profiles (Table~\ref{tab: sp_lines}). Effective rest wavelengths were taken from \citet{CLL77} and from \citet{Und94} for lines below and above 4800~\AA, respectively, to compute the radial velocities (RVs). Table 3 provides the journal of the observations and lists, for each spectral line, the measured RVs. 

The achieved RV accuracy is strongly dependent on the width of the lines, thus on the projected rotation rate of the star. Using a purely empirical approach, Fig.~\ref{fig: mu-sig} in this work and Fig.~10 in Paper~I provide us with the following guidelines. For slow rotators ($v \sin i < 50$~\kms), an rms accuracy of 1~\kms\ or better can be achieved. For  more typical rotation rates ($v \sin i \sim 100$-200~\kms), typical dispersions in the measured velocities are between 2 and 5~\kms. For fast rotators ($v \sin i > 300$~\kms), the precision of the measurements drops significantly, with rms dispersions close to 15~\kms.

 As in \citetalias{SGN08}, the spectral classification is based on the quantitative criteria of \citet{CA71}, \citet{Con73_teff}, \citet{Mat88} and \citet{Mat89}, that  rely on the EWs of given diagnostic lines. We adopt the following notations: $\log W'= \log W(\lambda4471) - \log W(\lambda4542)$,  $\log W''= \log W(\lambda4089) - \log W(\lambda4144)$ and  $\log W'''= \log W(\lambda4388) + \log W(\lambda4686)$, where the EWs are expressed in m\AA.  The first criterion, $\log W'$ can be used across the entire O-type range.  $\log W''$ and $\log W'''$ are restricted to stars strictly later that O6 and O8, respectively, while $\log W_{\lambda4686}$ can be used for O8 stars and earlier. The measurements and the corresponding spectral types are given in Table \ref{tab: EW} and discussed in Sect.~\ref{sect: ostar}.  Measured EWs for binary components only provide a lower limit on the real strength of the lines. This affects the latter two criteria, so that the values quoted in Table \ref{tab: EW} for $\log W'''$ and $\log W_{\lambda4686}$ are only lower limits.


\section{O-type stars in NGC\,6611} \label{sect: ostar}


\begin{figure}
\centering
\includegraphics[width=\columnwidth]{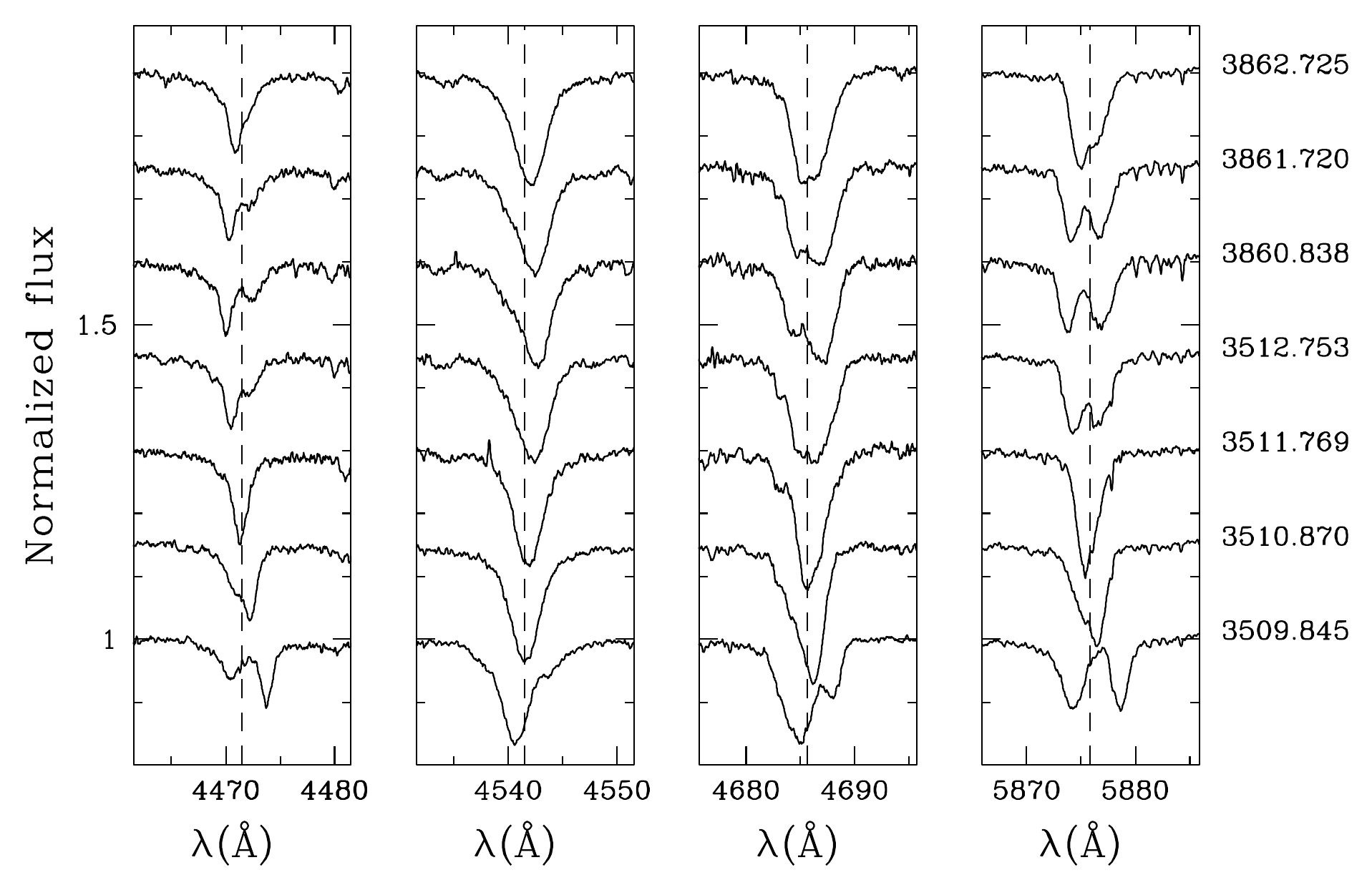}
\caption{{\bf \bd4923:} \hea\ll4471, 5876 and \heb\ll4542, 4686 line profiles at different epochs. The heliocentric Julian dates at mid-exposure are given at right-hand in format HJD$-$2\,450\,000. The spectra have been shifted along the $y$-axis for clarity. The vertical dashed lines indicate the adopted rest-wavelength for each spectral line. }
\label{fig: bd4923}
\end{figure}

\begin{table}
 \centering
 \begin{minipage}{80mm}
  \caption{{\bf \bd4923:} best-fit orbital solutions using the data set from the present work alone or combined with the \citeauthor{BMN99} RV measurements (primary only). $T$  (in HJD$-$2\,450\,000) is the time of periastron passage and is adopted as phase $\phi=0.0$ in Fig.~\ref{fig: bd4923os}. Quoted uncertainties correspond to 1-\s\ error-bars.}
\label{tab: bd4923os}
  \begin{tabular}{@{}lll@{}}
  \hline
Parameter              & This work            & Combined     \\
  \hline
$P$ (d)                & $ 13.2722 \pm 0.0027$ & $ 13.2677 \pm 0.0007 $ \\
$e$                    & $   0.285 \pm 0.030 $ & $   0.302 \pm 0.045   $ \\
\w\ (\degr)            & $   199.3 \pm 9.5   $ & $   188.5 \pm 9.8     $ \\
$T$                    & $2995.491 \pm 10.144$ & $3005.016 \pm 0.376   $ \\
$\gamma_1$ (\kms)      & $    30.7 \pm 4.3   $ & $    28.7 \pm 4.6     $ \\
$\gamma_2$ (\kms)      & $     3.4 \pm 5.3   $ & $     7.5 \pm 6.0     $ \\
$K_1$ (\kms)           & $    81.0 \pm 5.3   $ & $    82.6 \pm 7.0     $ \\
$K_2$ (\kms)           & $   140.2 \pm 9.2   $ & $   142.8 \pm 12.2    $ \\
$q=M_2/M_1$            & $   0.578 \pm 0.043 $ & $   0.578 \pm 0.043   $ \\
$M_1 \sin^3 i$ (\msol) & $     8.3 \pm 1.5   $ & $     8.6 \pm 2.1     $ \\
$M_2 \sin^3 i$ (\msol) & $     4.8 \pm 0.8   $ & $     5.0 \pm 1.2     $ \\
rms (\kms)             & $     7.9           $ & $    12.9             $ \\
\hline
\end{tabular}
\end{minipage}
\end{table}

\begin{figure}
\centering
Periodograms\\
\includegraphics[width=.49\columnwidth]{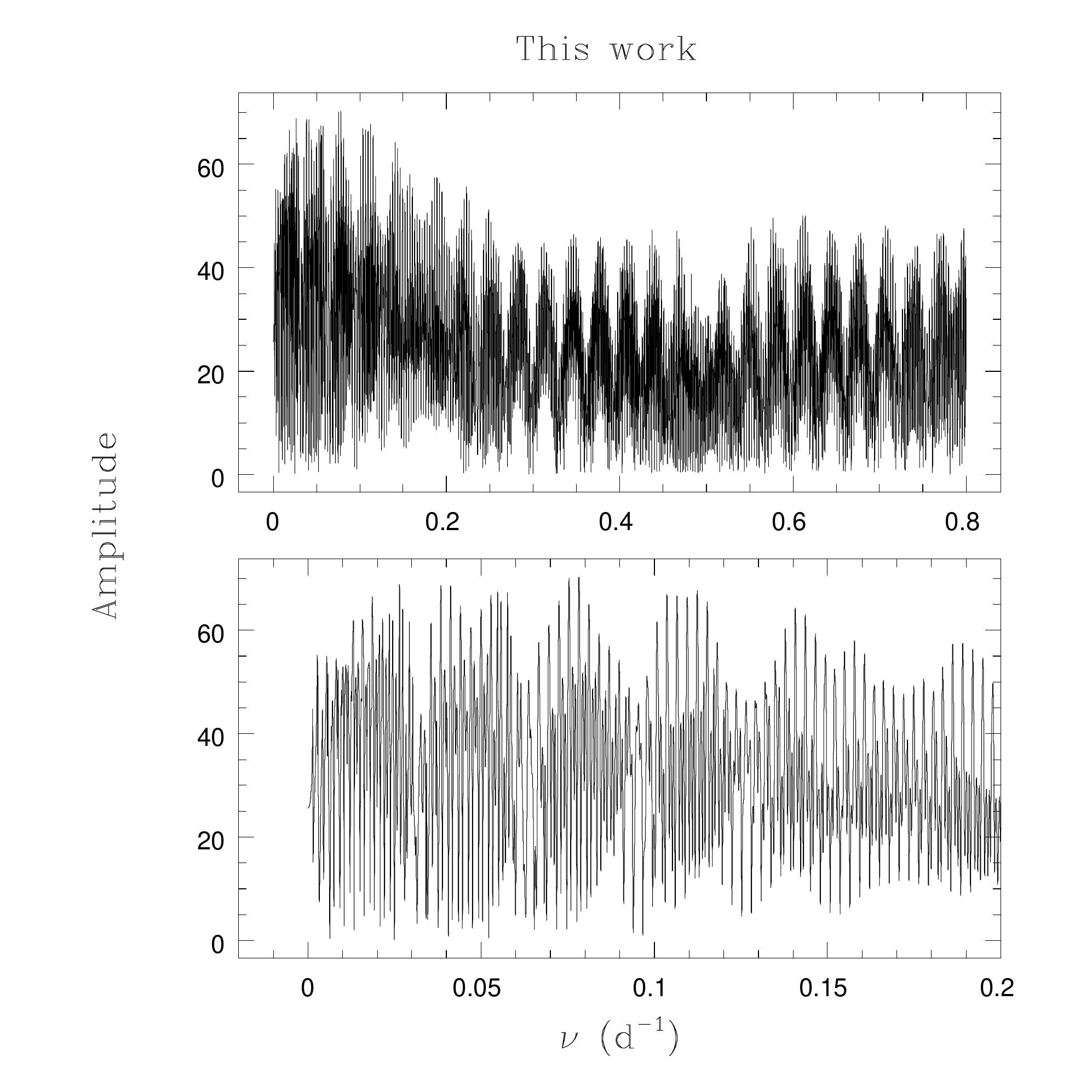}
\includegraphics[width=.49\columnwidth]{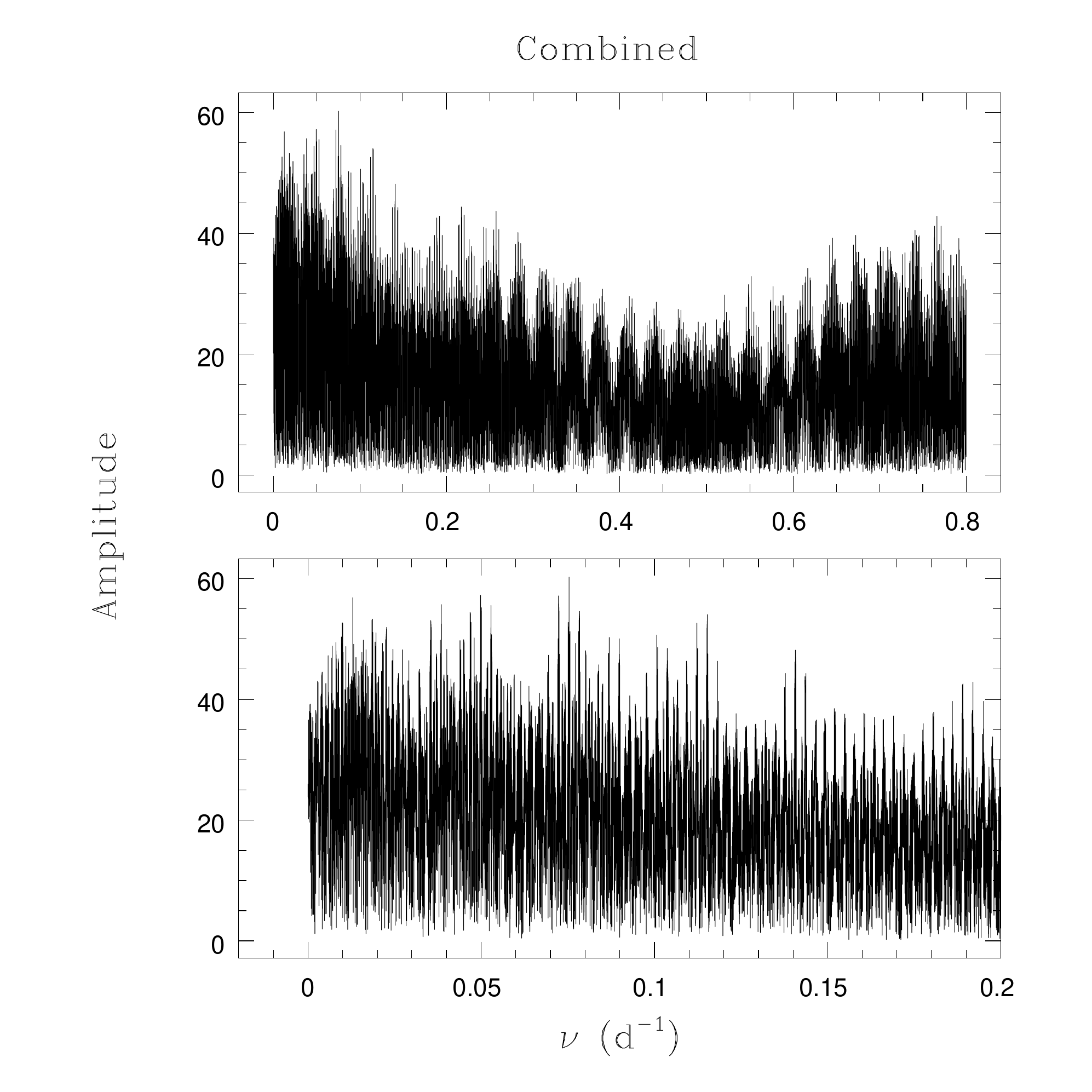}\\
Spectral windows\\
\includegraphics[width=.49\columnwidth]{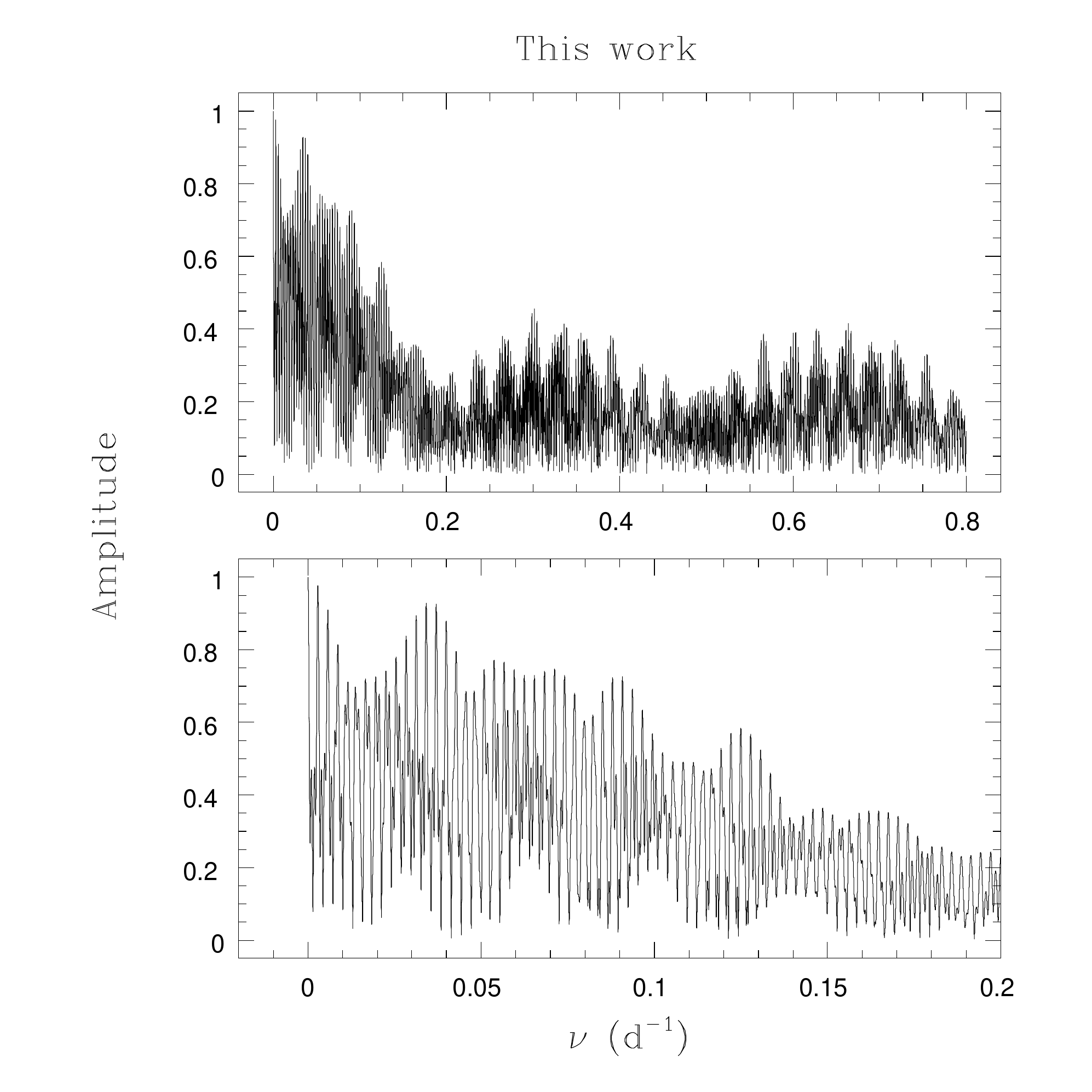}
\includegraphics[width=.49\columnwidth]{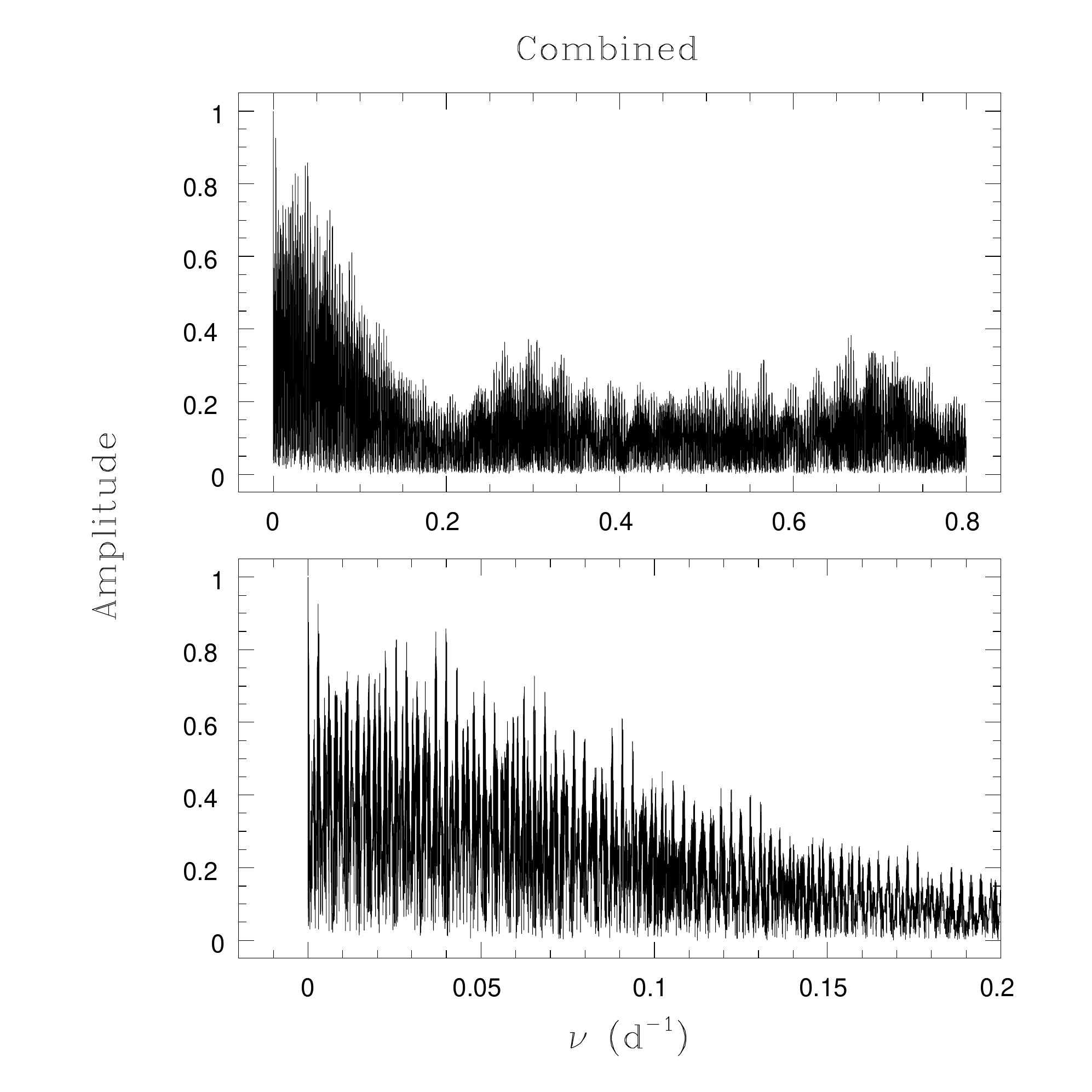}
\caption{{\bf \bd4923:} periodograms (upper row) and spectral windows (lower row) computed using the data from this work alone (left-hand column) or combined with the \citeauthor{BMN99} measurements (right-hand column). }
\label{fig: bd4923hmm}
\end{figure}
\begin{figure}
\centering
\includegraphics[width=.49\columnwidth]{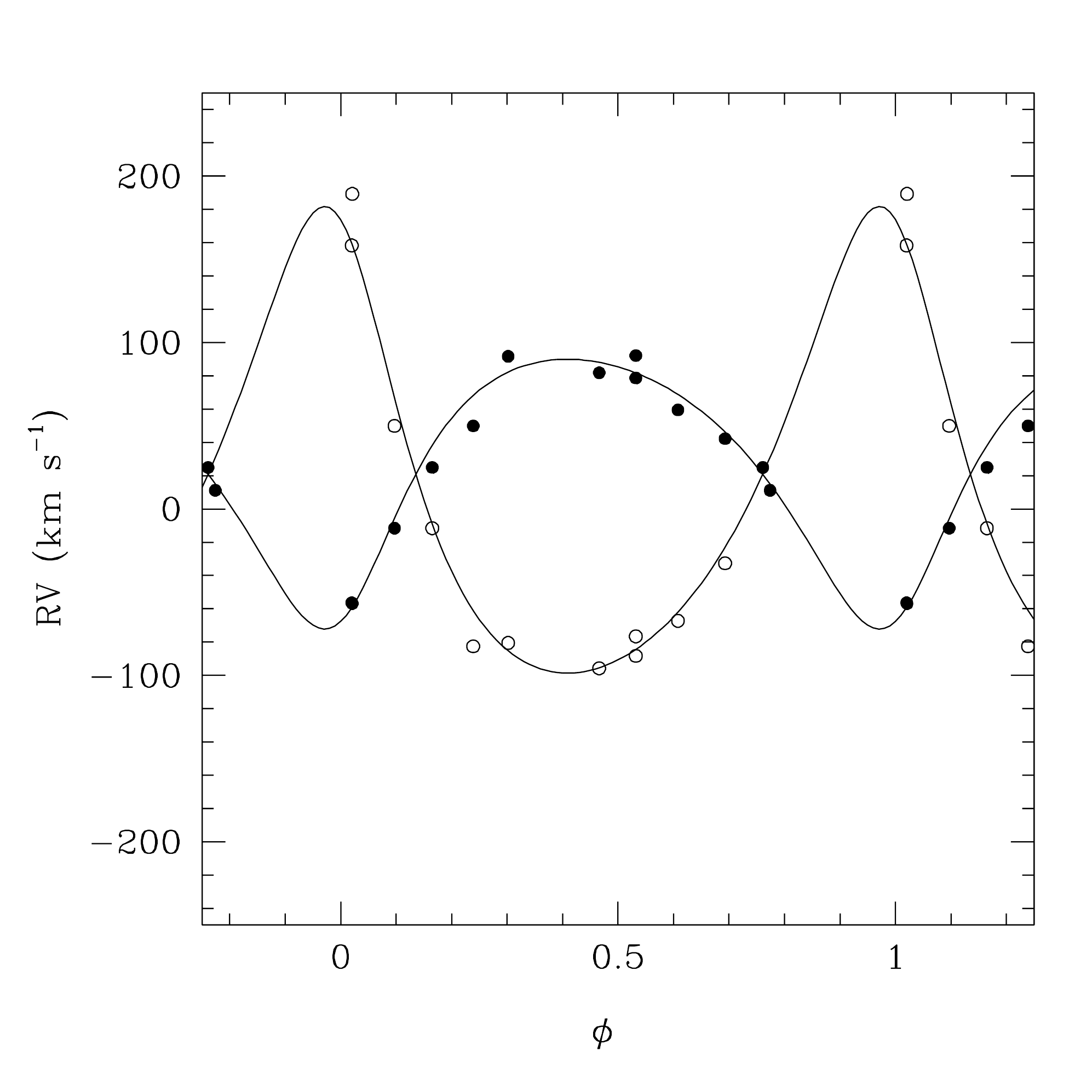}
\includegraphics[width=.49\columnwidth]{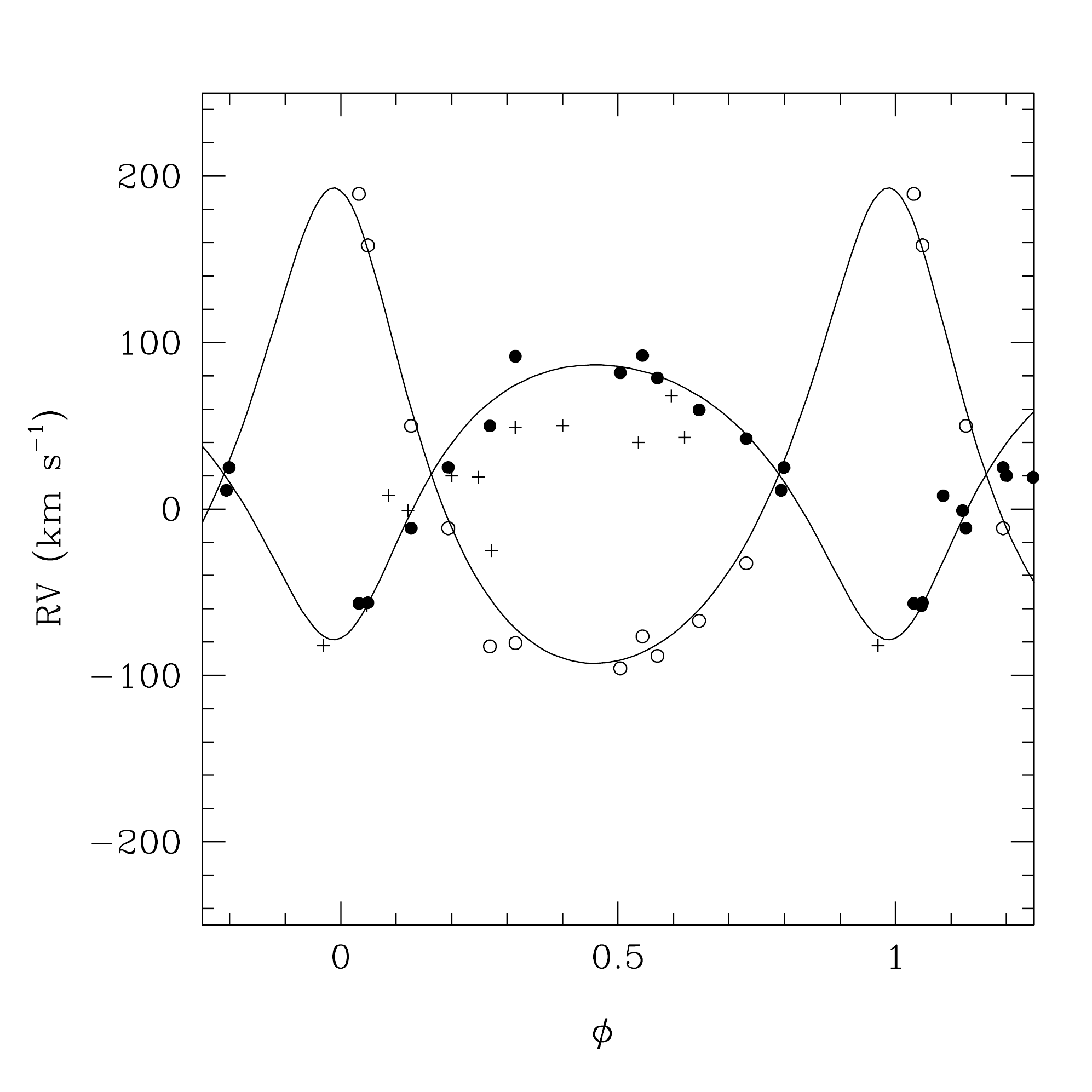}
\caption{{\bf \bd4923:} RV-curves corresponding to the orbital solutions of Table~\ref{tab: bd4923os}. {\it Left: } using data from this work. {\it Right: } combined with the \citeauthor{BMN99} measurements. Filled and open symbols indicate the primary and secondary RV measurements from this work while crosses give the primary RVs from \citeauthor{BMN99}}
\label{fig: bd4923os}
\end{figure}


This section briefly discusses the data associated with each O-type object in our sample, with 
a particular emphasis on the spectral properties and on the multiplicity status.  As mentioned earlier, 
\ngc\ hosts 13 O-type stars. Table~\ref{tab: ID} provides the cross-identification between HD/BD numbers and the numeration of \citet{Wal61}, \citet[][ \citetalias{ESL05mnras}]{ESL05mnras} and \citet[][ \citetalias{GPM07}]{GPM07}. 
It also lists the stellar parameters obtained by \citet{DSL06} from atmosphere model fitting. The four faintest O-stars were not re-observed 
as part of our monitoring program, so that only one or two spectra are available for each object. Those four stars are thus omitted in the rest of this paper. 

This section is organised as follows. We first present the gravitationally-bound systems (Sect.~\ref{ssect: gbs}).
In Sect.~\ref{ssect: comp}, we discuss objects whose spectra most likely present multiple signatures,
although our current data set does not allow us to conclude whether the different components are 
gravitationally-bound or arise from spurious alignment. Finally, Sect.~\ref{ssect: sgl} summarizes
the properties of the remaining, presumably single, O-type stars.

\subsection{Gravitationally-bound systems}\label{ssect: gbs}

\sss{\bd4923 (W175)}
Classified as O5~V((f$^*$)) by \hms,  BD$-$13\degr4923 was first suspected to be an O+O binary with a period larger than eight days by \bmn.  Based on the fact that the \hea\ and \heb\ lines were moving in the opposite direction, these authors suggested that the primary was an early O star and the secondary, a late one, a fact later confirmed by \esl. \dse\ further reported a low-mass visual companion at 0.67\arcsec ($\approx$1200~A.U.), with an estimated mass between 2 and 4~\msol.


We collected 10 additional spectra over three years. At least four of them clearly show the signatures of the two components both in the \hea\ and \heb\ lines (Fig.~\ref{fig: bd4923}), allowing us to put additional constraints on the orbital and physical properties of the components. Using the well-separated spectra only, we refined the spectral classification to O4~V((f))+O7.5~V with, respectively, the O5 and O7 spectral sub-type at 1-\s.  The third, more distant, component is likely a late-B/early-A star. It is thus expected to contribute to less than 1\%\ of the total continuum level and will not affect the RV and EW measurements of the two O stars. It has thus been neglected in the present analysis.

Because we only have a limited data set, we also made use of previously discarded data. First,  one of our \feros\ spectra was obtained late during morning twilight and is contaminated by the solar spectrum. From \heb\l4686 and redward,  the solar and stellar spectra can however be easily separated, allowing us to recover most of the prominent stellar lines in that region. Second, the WHT-\isis\ spectrum from \citet{ESL05mnras} exhibits a clear wavelength calibration shift in the \l4800~\AA\ setting.  To solve this issue, we have cross-calibrated the \isis\ wavelength solution with respect to our own data, by using the narrow DIBs at \ll4763, 4780 and 4964 \citep{HYS08}. This allowed us to achieve a corrected wavelength calibration with an rms dispersion better than 0.1~\AA\ ($\sim$5~\kms), sufficient thus for our purpose. 

In the following, we focus explicitly on the \heb\l4686 line which is the only strong SB2 line common to all our spectra. To improve the disentangling of heavily blended profiles, we adopted the line shapes as determined on widely separated spectra obtained during the same run, thus fitting the Doppler shifts only. Although it is not clear whether our data cover any of the  extrema of the RV-curve, we attempted to constrain the orbital period using the Fourier method of \citet{HMM85}, as revisited by \citet{GRR01}. The obtained periodogram shows two dominant peaks at $\nu\approx0.075$ and $0.078$~d$^{-1}$ (Fig.~\ref{fig: bd4923hmm}). Including the \heb\ measurements of \bmn\ further allows us to determine that the signal at  $\nu\approx0.075$~d$^{-1}$ is most likely to correspond to the true periodicity of the orbit. 

Using the corresponding period $P\approx13.2679$~d as a first guess, we used the Li\`ege Orbital Solution Package \citep[LOSP, ][]{SaG09} to compute a preliminary SB2 orbital solution. Table~\ref{tab: bd4923os} and Fig.~\ref{fig: bd4923os} respectively give the best-fit orbital parameters and the corresponding RV-curves. We note that the period uncertainties given in Table~\ref{tab: bd4923os} only refer to the fitted  models. As such, they do not account for the presence of multiple peaks in the periodogram. Both solutions, computed with and without including literature data, are in good agreement although Fig.~\ref{fig: bd4923os} reveals some systematic deviations between the best fit solution and the RV points of \bmn. Yet, the latter measurements might suffer from significant systematic errors because the SB2 signature was not separated. With $e\approx0.3$, the obtained solution is significantly eccentric. The derived mass-ratio is in perfect agreement with the estimated spectral types of the components. Given the computed minimal masses, one would predict an orbital inclination about 30 to 40\degr, so that no eclipses are expected.

\begin{figure}
\centering
\includegraphics[width=\columnwidth]{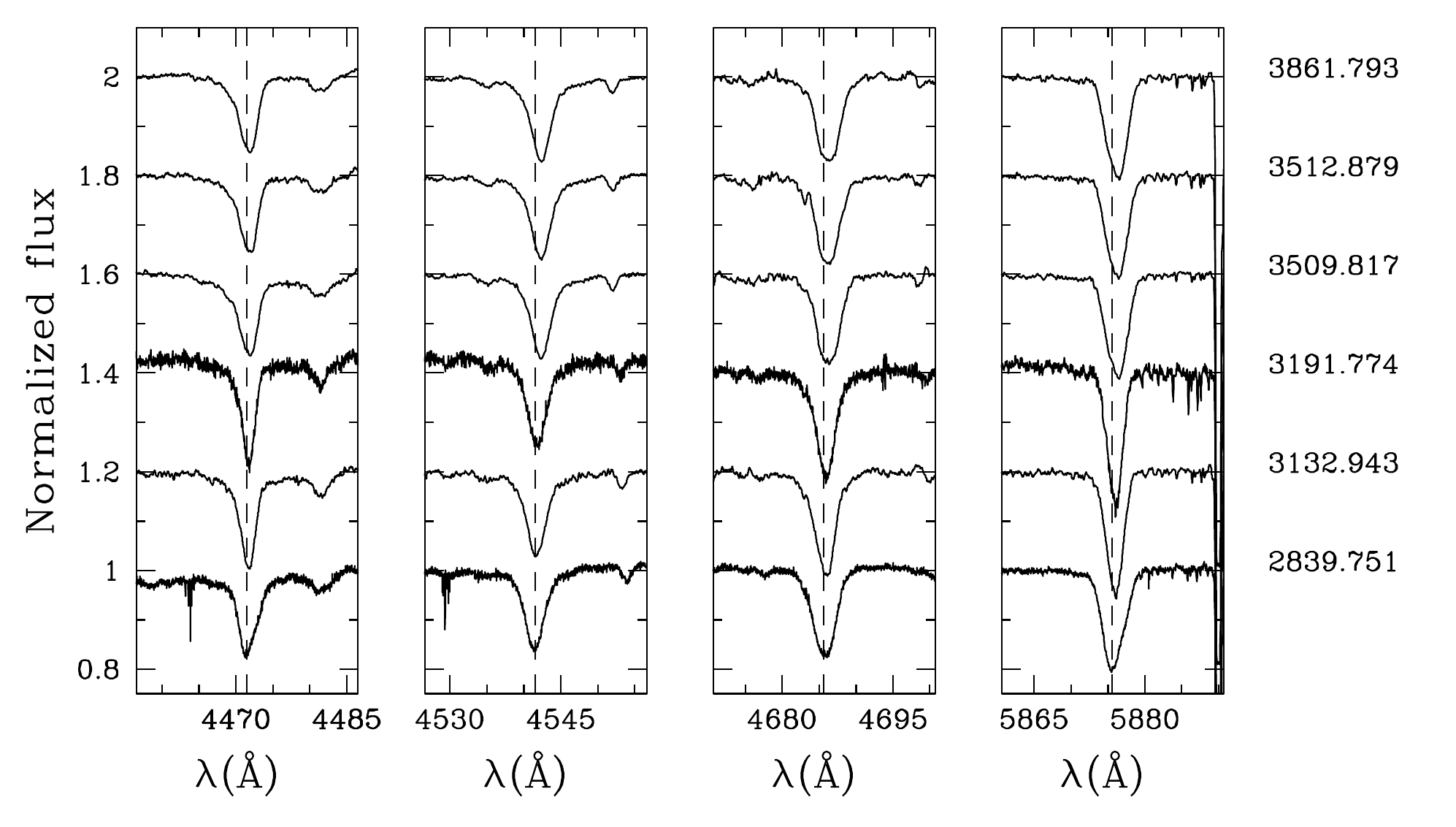}
\caption{{\bf HD\,168075:} \hea\ll4471, 5876 and  \heb\ll4542, 4686 line profiles at different epochs. Note the changing asymmetry of the \hea\ lines.}
\label{fig: hd075}
\end{figure}

\sss{HD\,168075 (W197)}
So far classified as an O6-O7~V((f)) star \citep{ESL05mnras}, HD\,168075 was first suggested to be a binary by \citet{CLL77}. Later on, \bmn\ reported no short-term variations but mentioned significant changes in data taken two years apart. These authors further reported the detection of \sic, \nc\ and \ob\ absorption lines normally not present in a typical O7 spectrum. 

Our data sample, formed by about 10 spectra collected over three years, confirms this object is a long period binary. As found by \citeauthor{BMN99}, we clearly detect \sic, \nc\ and \ob\ lines in absorption, whose motion is  anti-correlated with the \hea\ and \heb\ line motion, thus confirming their association with the secondary star. As observed in Fig.~\ref{fig: hd075}, the \hea\l4471 and \heb\l4686 lines display clear profile changes, likely due to the blending of the primary and secondary signatures, while the \heb\l4542 line profile seems to remain constant in shape. This strongly suggests the secondary to be a B0-B1 star. Using the \hea\l4471 over \heb\l4542 EW ratio, we adopt a O6.5~V((f)) spectral type for the primary, with a O7 sub-type well within 1-\s.  We finally note that the \hea\l4471 line EW might have been overestimated because of the secondary blend, so that the primary could be slightly hotter than the adopted spectral sub-type.

 Our data alone do not allow us to definitely constrain the orbital properties of HD\,168075. Using a much larger data set, \citet{BarbaVina} recently proposed a preliminary orbital solution, with a period of 43.6~d and a slightly eccentric orbit $e=0.17$. They independently confirmed the O6.5~V((f)) classification for the primary and proposed that the secondary is a B0.2~III star.

\begin{figure}
\centering
\includegraphics[width=\columnwidth]{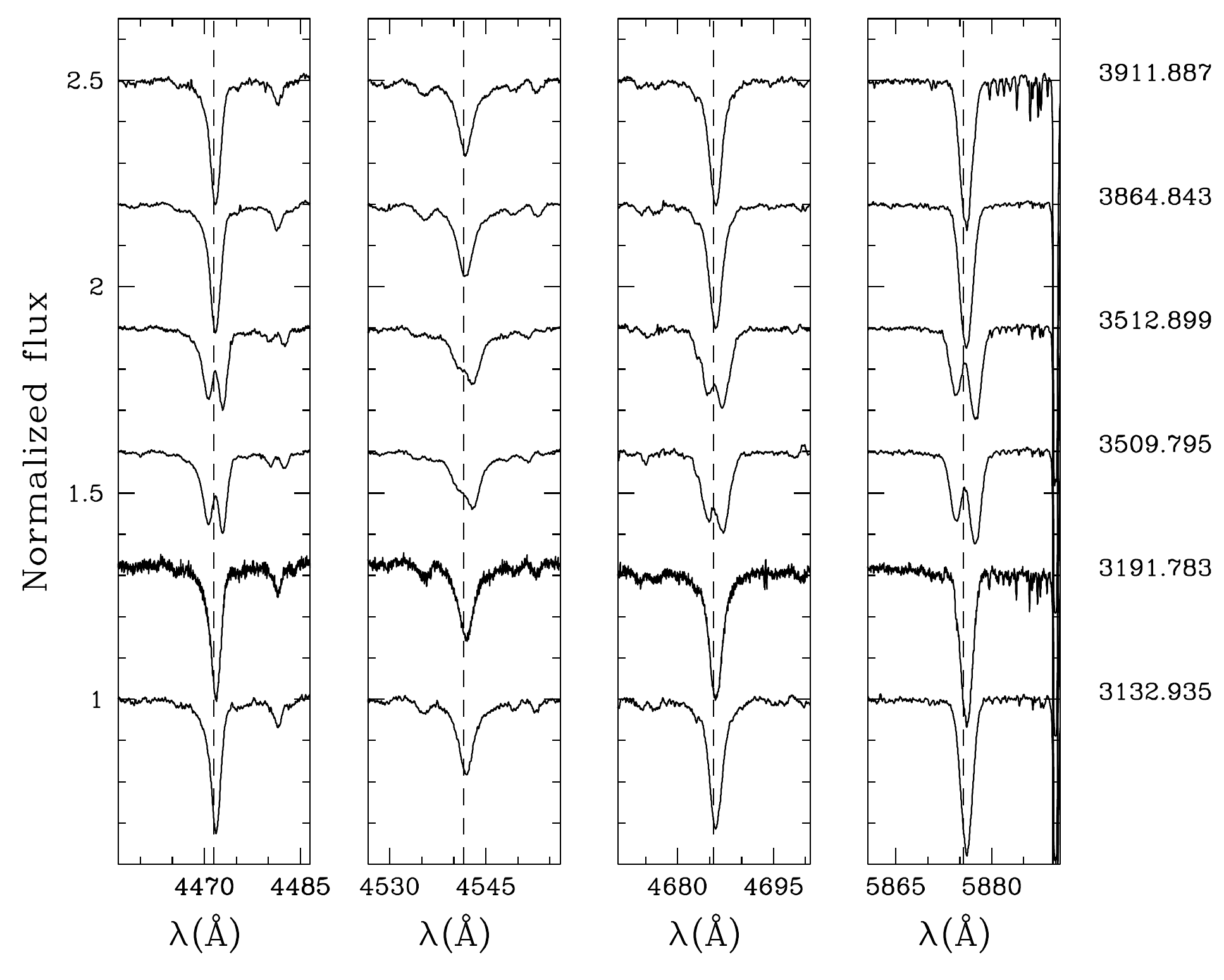}
\caption{{\bf HD\,168137:} \hea\ll4471, 5876 and  \heb\ll4542, 4686 line profiles at different epochs. Note the clear SB2 signature at HJD$-$2\,450\,000 $\sim$ 3509.8 and 3512.9.}
\label{fig: hd137}
\end{figure}

\sss{HD\,168137 (W401)}
HD\,168137 is known as an O8.5~V isolated star. Between May 2004 and May 2006, we acquired 10 \feros\ spectra and, in March 2009, we obtained one additional \feros\ observation. Our prime data set is completed by one \gir\ spectrum and one \uves\ spectrum from \esl\ obtained in 2003. While our May 2004, 2006 and 2009 observations are very similar to the spectra of \citeauthor{ESL05mnras}, the May 2005 data reveal a clear SB2 signature (Fig.~\ref{fig: hd137}). No further RV variations could be detected within the 2005 campaign. Similarly, no variations are seen between the May and June 2006 spectra, suggesting a period larger than 30 days. Using the well-separated spectra only, we revise the spectral classification to O7~V+O8~V, with the O6.5 and O7.5 subclasses at 1-\s\ respectively for the primary and secondary components. A long term monitoring is required to bring further constraints on the orbital properties of this system.

\begin{figure}
\centering
\includegraphics[width=0.7\columnwidth]{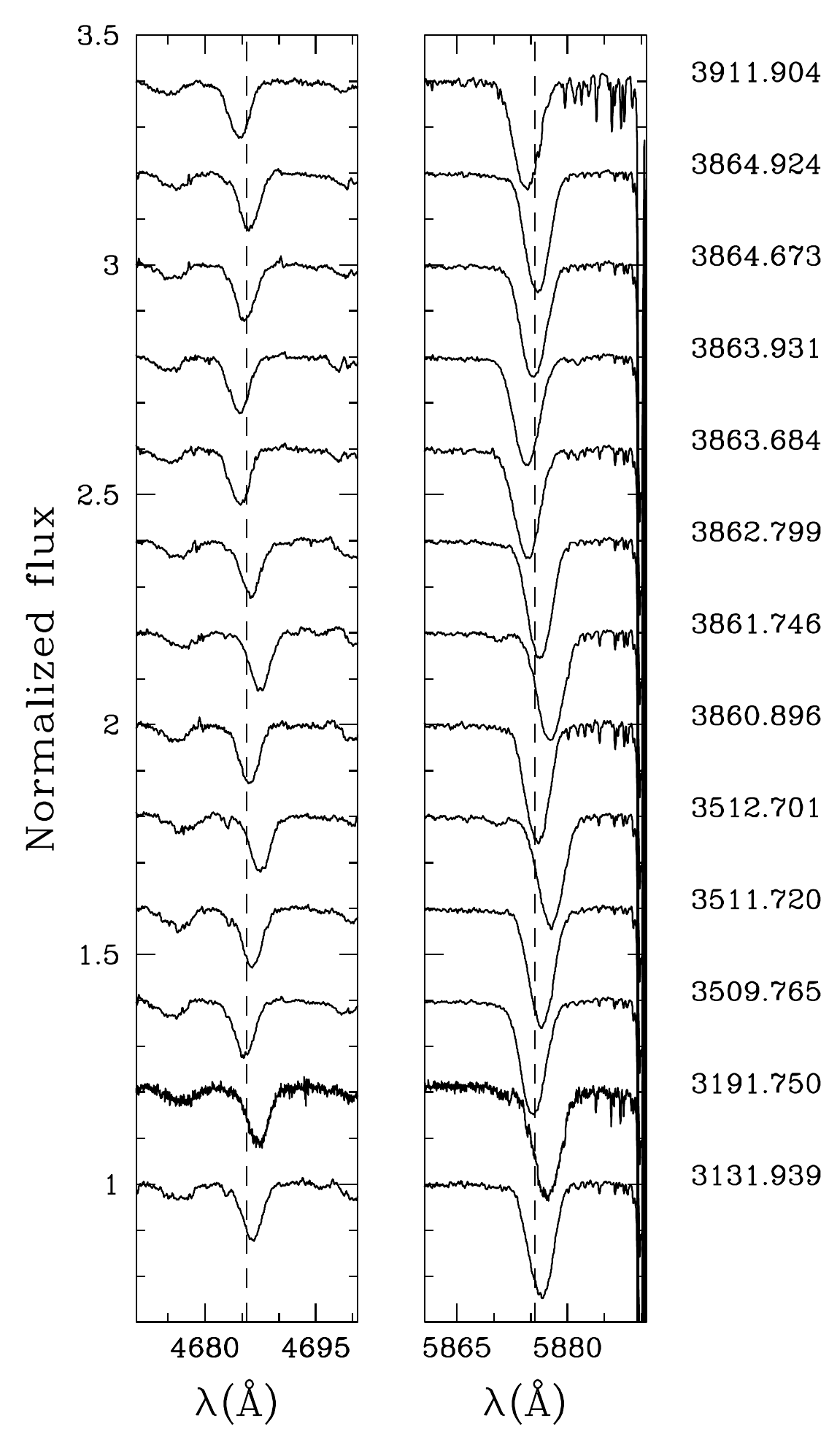}
\caption{{\bf HD\,168183:}  \heb\l4686 and \heb\l5876 line profiles at different epochs. Note the faint SB2 signature at HJD$-$2\,450\,000 $\sim$ 3512.7 and 3861.7. }
\label{fig: hd183}
\end{figure}

\begin{figure}
\centering
Periodograms\\
\includegraphics[width=0.49\columnwidth]{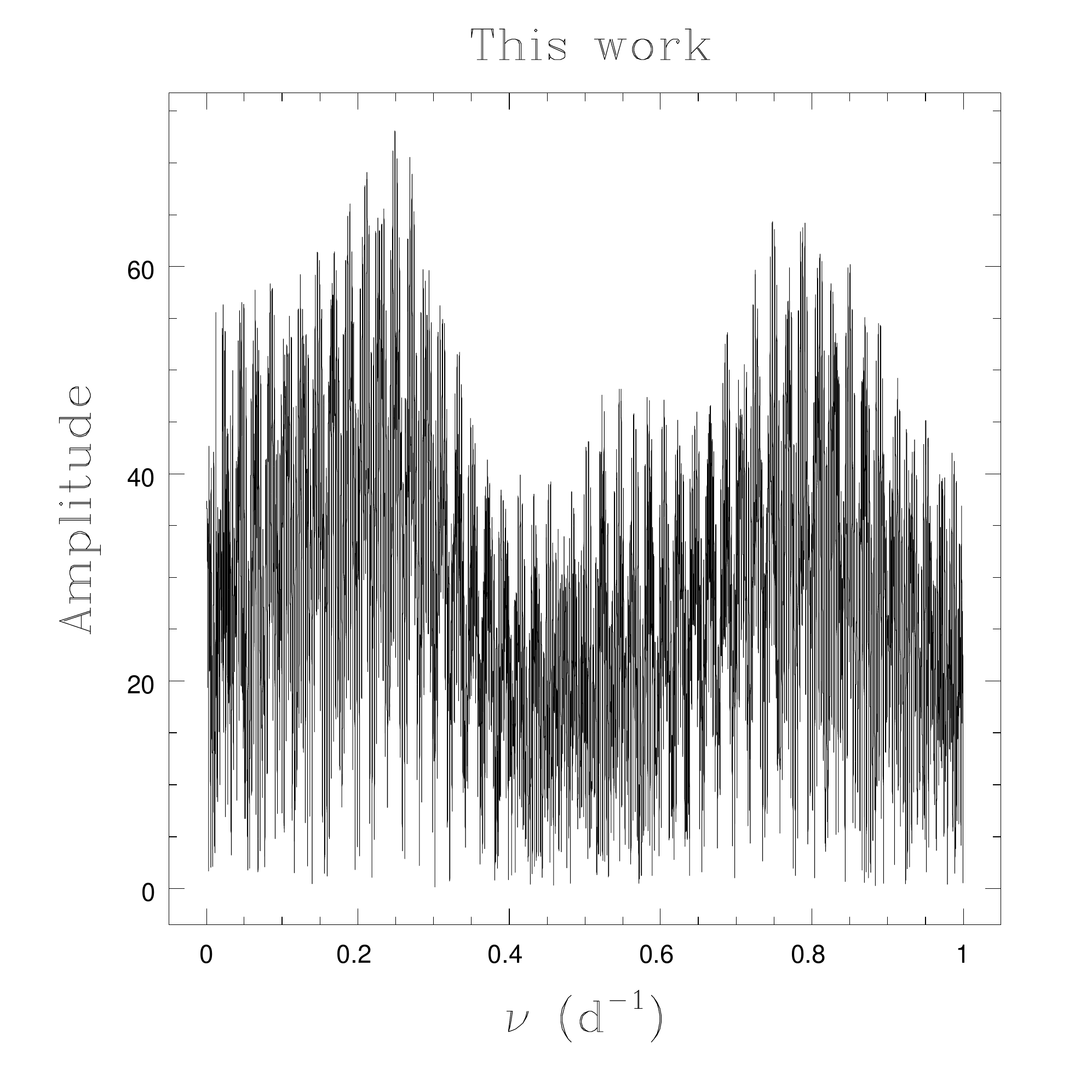}
\includegraphics[width=0.49\columnwidth]{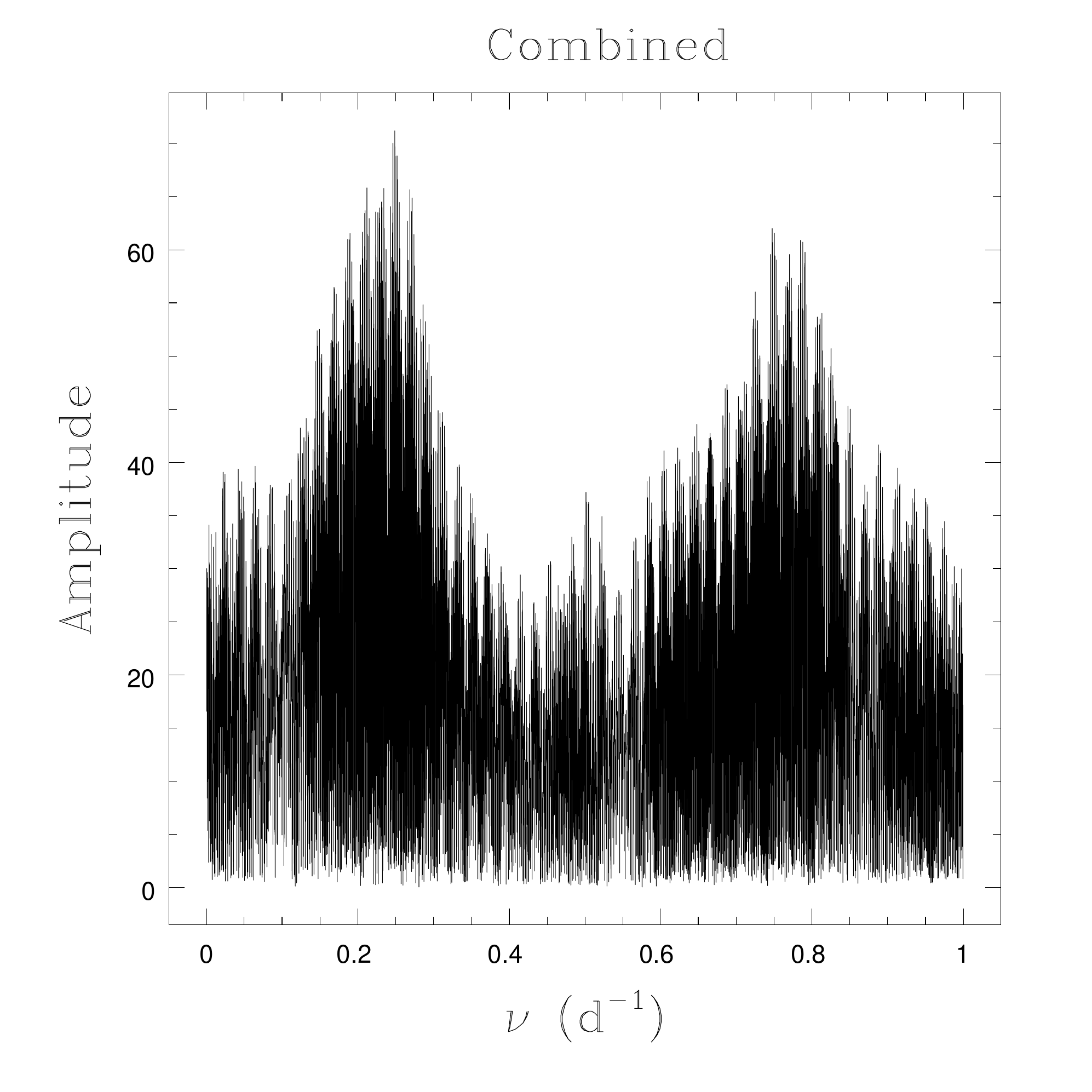}
Spectral windows\\
\includegraphics[width=0.49\columnwidth]{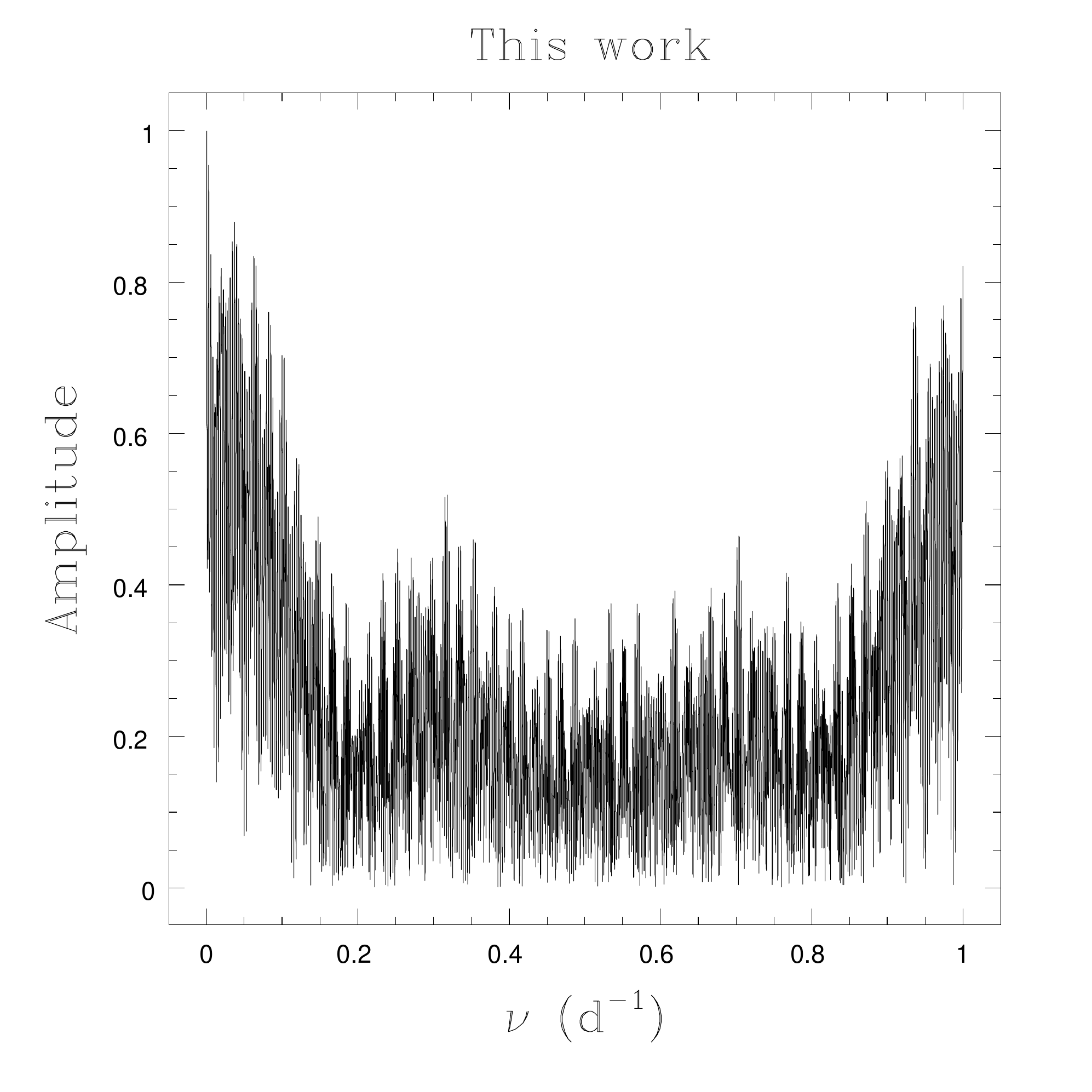}
\includegraphics[width=0.49\columnwidth]{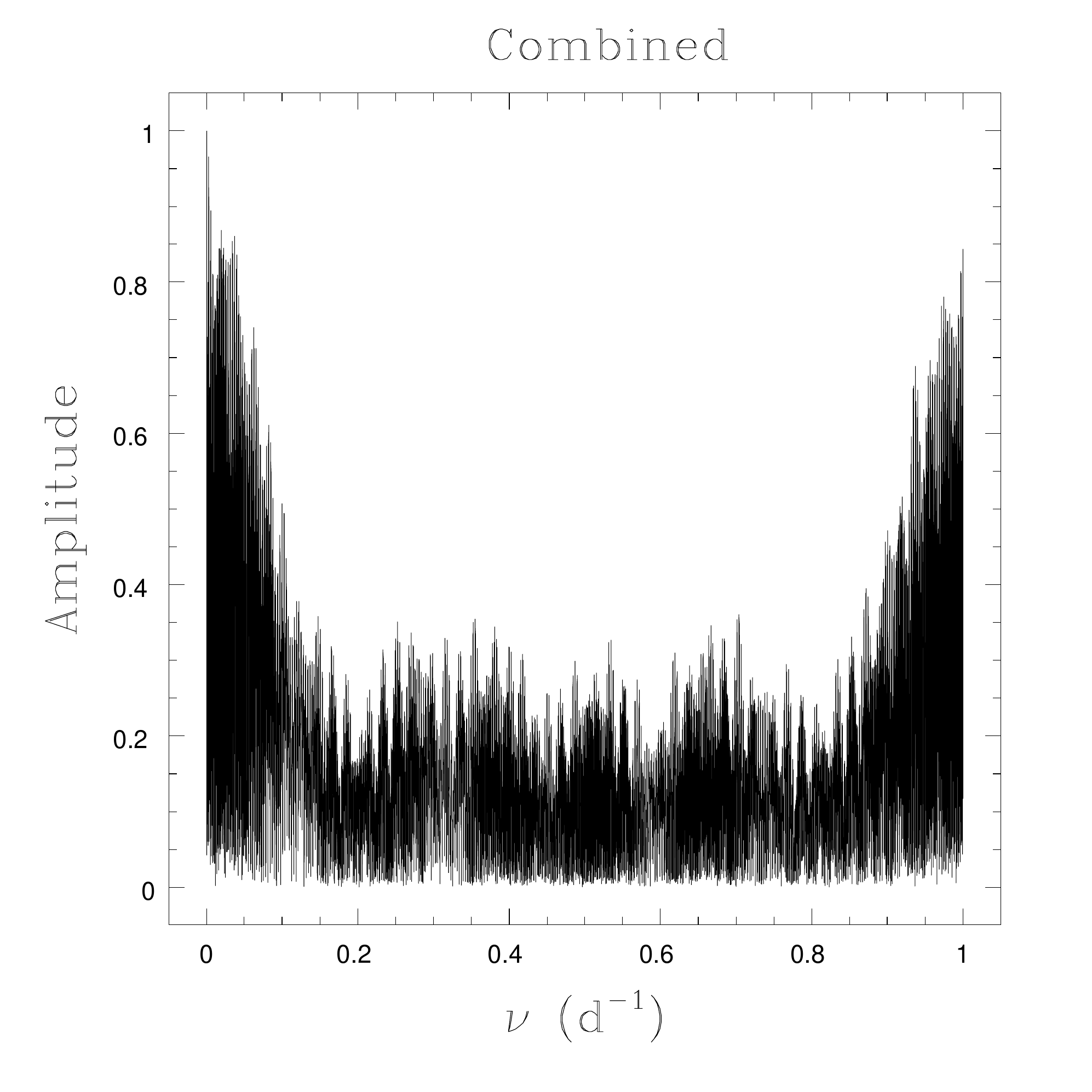}
\caption{{\bf HD\,168183:} periodograms (upper row) and spectral windows (lower row) computed using the \hea\ data from this work alone (left-hand column) or combined with the \citeauthor{BMN99} measurements (right-hand column).}
\label{fig: hd183hmm}
\end{figure}

\begin{figure}
\centering
\includegraphics[width=0.49\columnwidth]{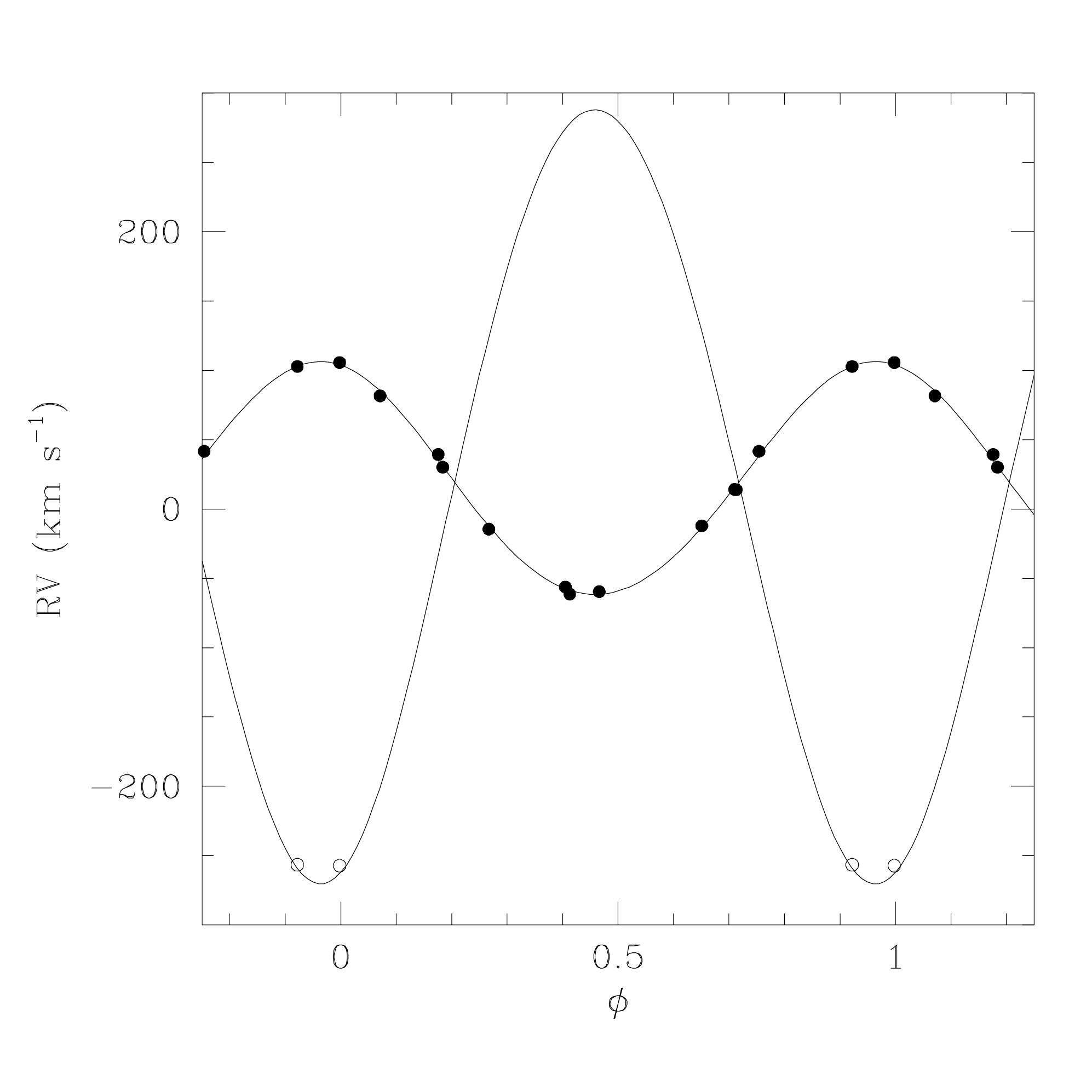}
\includegraphics[width=0.49\columnwidth]{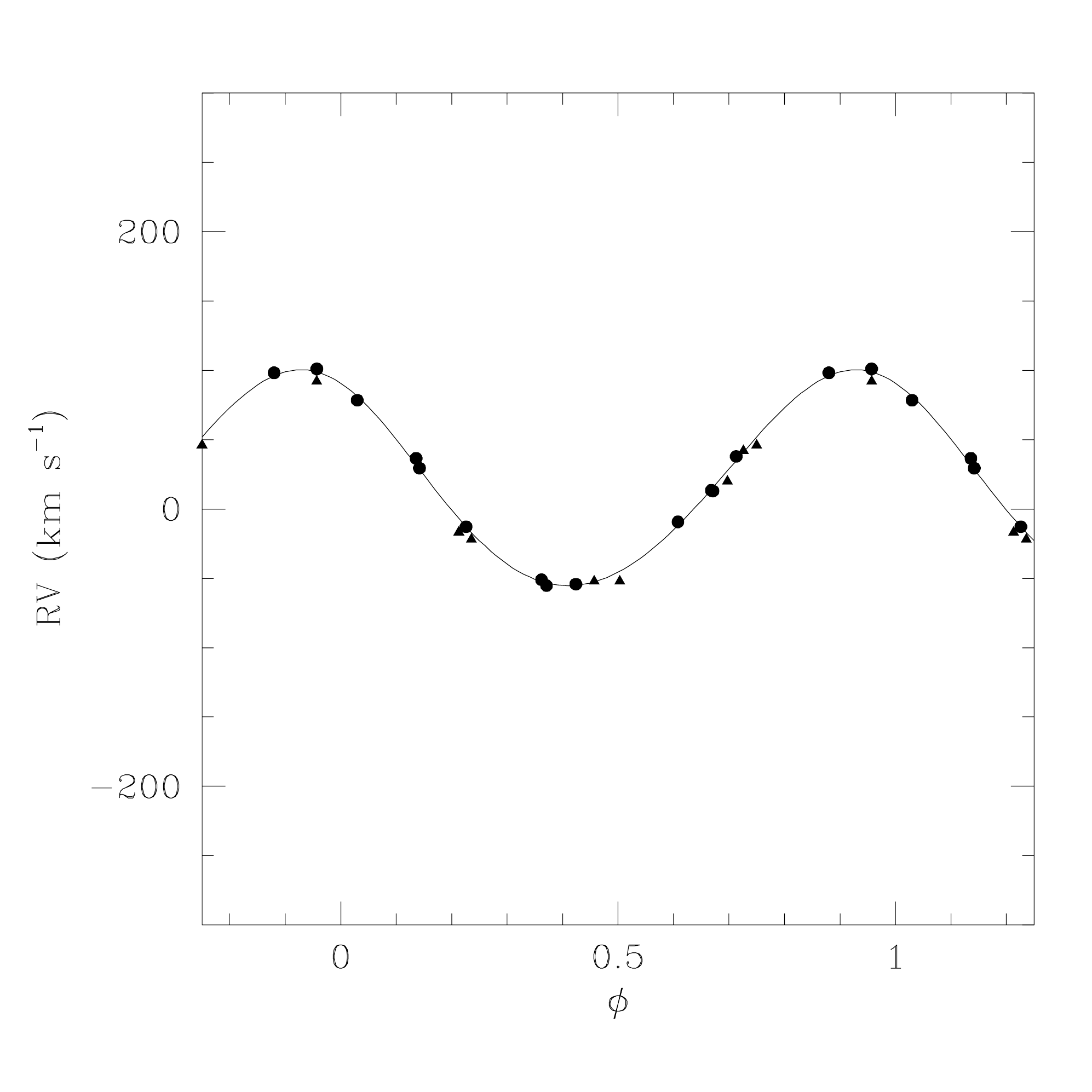}
\caption{{\bf HD\,168183:}  RV-curves corresponding to orbital solutions of Table~\ref{tab: hd183os}, based on the \hea\l5876 line data set (left) and on the average of the \hea\ line RVs combined with \citeauthor{BMN99} measurements (right). Filled and open circles show primary and secondary RV measurements from this work while filled triangles show the  \citeauthor{BMN99} data.}
\label{fig: hd183rvc}
\end{figure}

\begin{table*}
\centering
\caption{{\bf HD\,168183:} orbital solutions obtained using different data sets:  the RVs of the \hea\l5876 line (Col.~2), the average of the \hea\ measurements (Col.~3) and the average of the \hea\ measurements combined with  the  \citeauthor{BMN99} data (Col.~4). The usual notations have been used. $T$ (in HJD$-$2\,450\,000) is the time of periastron passage and is adopted as phase $\phi=0.0$ in Fig.~\ref{fig: hd183rvc}. Quoted uncertainties correspond to 1-\s\ error-bars.}
\label{tab: hd183os}
\begin{tabular}{llll}
\hline
                  & \hea\l5876 line &  $\overline{\mathrm{He\sc{I}~line}}$ & Combined \\
\hline
$P$ (d)           &   4.01554 $\pm$ 0.00017   &  4.01556 $\pm$ 0.00015 & 4.015575 $\pm$ 0.000022 \\ 
$e$               &     0.039 $\pm$ 0.015     &    0.052 $\pm$ 0.014   &    0.059 $\pm$ 0.015    \\
$\gamma$ (\kms)   &      19.2 $\pm$ 1.0       &     19.3 $\pm$ 0.9     &     18.5 $\pm$ 0.8      \\
$K$ (\kms)        &      83.9 $\pm$ 1.5       &     78.5 $\pm$ 1.3     &     77.8 $\pm$ 1.3      \\
$T$               &  2998.718 $\pm$ 0.388     & 2998.786 $\pm$ 0.244   & 2998.881 $\pm$ 0.195    \\
\w\ (\degr)       &      13.9 $\pm$ 35.8      &     20.1 $\pm$ 22.7    &     28.9 $\pm$ 17.5     \\
$f_\mathrm{mass}$  &     0.245 $\pm$ 0.013     &    0.200 $\pm$ 0.010   &    0.195 $\pm$ 0.009    \\ 
rms (\kms)        &       3.3                 &      2.8               &      3.7                \\       
 \hline
\end{tabular}
\end{table*} 

\sss{HD\,168183 (W412)}
At $\sim$13\arcmin\ SE of  the center of the cluster, HD\,168183 is a known SB1 eclipsing binary with a period close to 4~d \citep{Hip97, BMN99}. The object has been classified O9.5~I by \citet{HMS93} and \citet{BMN99}, and B0~III by \citet{ESL05mnras}. With a significant \heb\l4542 line in absorption, HD\,168183 cannot be a B star. Indeed the EW ratio of \hea\l4471 to \heb\l4542 points to an O9.5 spectral type. With \hea\l4144 only slightly shallower than \sid\l4089, plus \heb\l4686 and \halp\ in absorption, the properties of the spectrum of HD\,168183 are definitely not those of a supergiant. Given the brightness of HD\,168183, we finally adopt a O9.5~III spectral classification for the primary.

Thanks to their high SNR, our data reveal the secondary signature for the first time. Though faint, the secondary spectrum is indeed clearly visible in the \hea\l5876 line, as a broad and very shallow component, peaking at $\approx$2\% of the continuum level (Fig.~\ref{fig: hd183}). Because of the large difference in the line EWs (a factor $\sim$20 for \hea\l5876), indicating a significant difference in flux, and because we could not detect the secondary signature in the \heb\ lines, we suggest that the secondary is a mid-B type star. 

As for \bd4923, we used the Fourier technique of \citet{HMM85} and \citet{GRR01} to estimate the orbital period of the system. Resulting periodograms are shown in Fig.~\ref{fig: hd183hmm} and present a clear peak at $\nu\approx0.25$~d$^{-1}$. Using LOSP, we first computed the primary orbital solution. Significantly better residuals are obtained using a limited eccentricity compared to a circular solution. We derived a RV-curve semi-amplitude $K_1$ and a mass function $f_\mathrm{mass}$ larger than those proposed by \bmn. Combining our data with theirs yields essentially the same best-fit solution (Table~\ref{tab: hd183os}).

Assuming that the primary orbital solution is correct and that both components share the same systemic velocity, it is  in principle possible to constrain the secondary orbit with a single RV measurement. For the purpose of this exercise, we adopt the \hea\l5876 orbital solution of Table~\ref{tab: hd183os}, propagating the uncertainties on the primary parameters at the first order by means of the theory of error propagation \citep{Bev69}. Using the two observations where the SB2 signature is well seen, we estimate $K_2=279.1\pm10.9$~\kms. This yields $q=M_2/M_1=0.301\pm0.013$,  $M_1 \sin^3 i=15.3\pm1.9$~\msol\ and $M_2 \sin^3 i=4.6\pm0.4$~\msol. The corresponding RV-curves are displayed in Fig.~\ref{fig: hd183rvc}. While this result supposes that the adopted time series alias corresponds to the true period, we note that the obtained mass ratio is in good agreement with the component spectral types as previously discussed. Under such assumption, the orbital inclination would be $i\approx65$\degr, compatible with the presence of eclipses as detected by \hipparcos.

\subsection{Composite systems}\label{ssect: comp}

\begin{figure}
\centering
\includegraphics[width=\columnwidth]{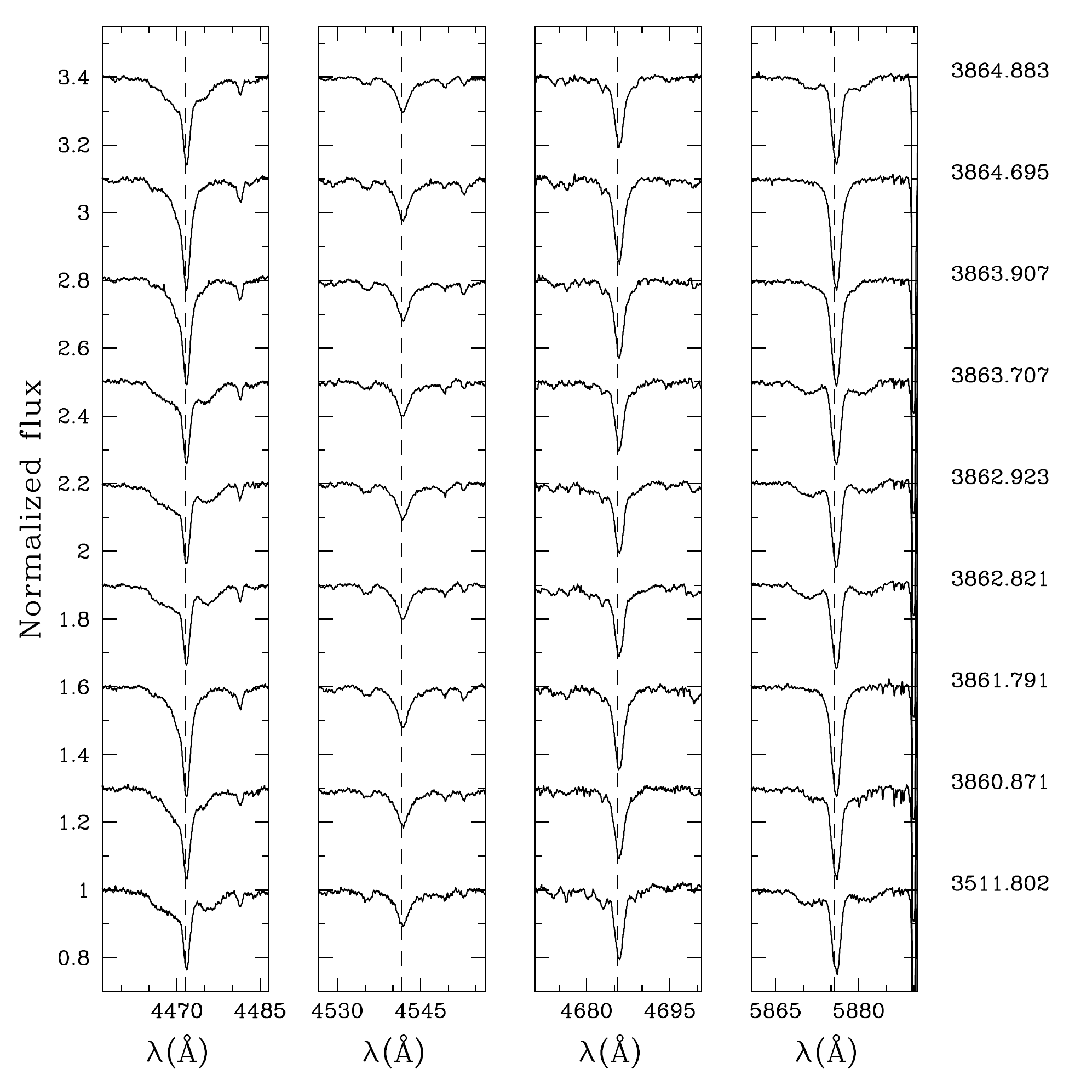}
\caption{{\bf \bd4929:} \hea\ll4471, 5876 and  \heb\ll4542, 4686 line profiles at different epochs. The close B+B pair signature is clearly visible in the \hea\ lines at  HJD$-$2\,450\,000  from 3862.8 to 3863.7. }
\label{fig: bd4929}
\end{figure}

\begin{figure*}
\centering
\includegraphics[width=15cm]{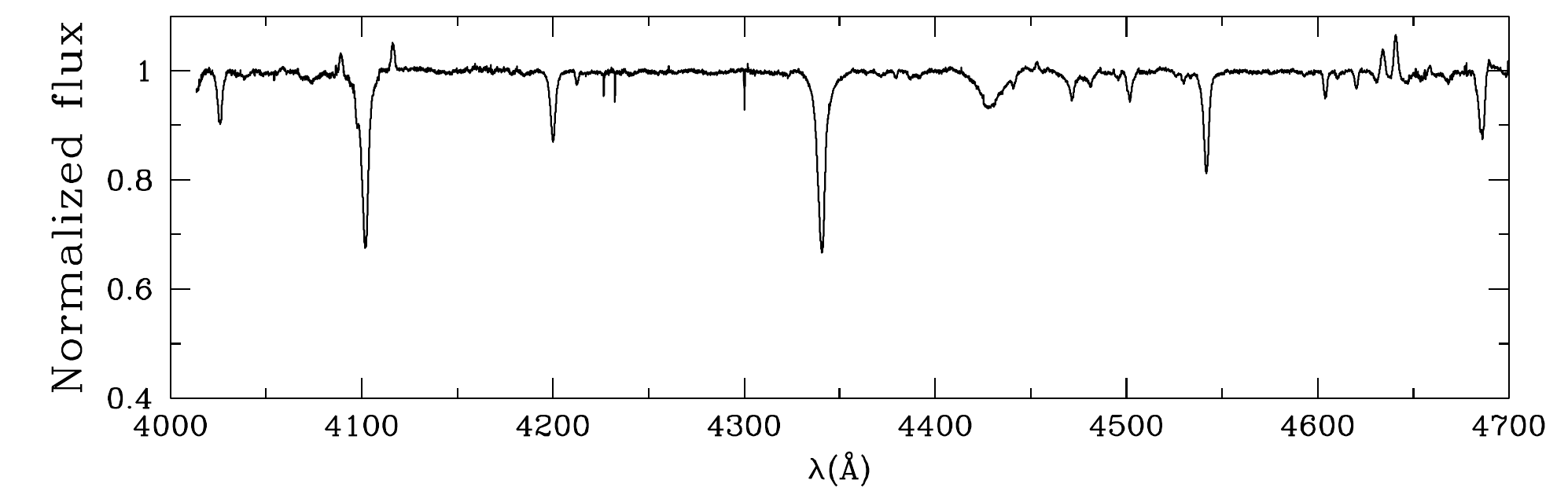}
\caption{{\bf HD\,168076:} spectrum in the range 4000-4700~\AA. Note the \sid\ll4089-4116 and \nc\ll4634-4640-4642 lines in emission and the \ne\ll4604-4620 absorption. }
\label{fig: hd076spec}
\end{figure*}

\sss{\bd4929 (W314)}
Classified B0~V by \hms\ and \bmn, the star's spectral type was revised to O9~V by \esl. The latter authors further noticed slight profile asymmetries, suggesting  that BD$-$13\degr4929 is an SB2 binary system. Our new data allow us to shed new light on the nature of \bd4929. From Fig.~\ref{fig: bd4929}, it is obvious that \bd4929 is an SB3 system, formed by a short-period binary with nearly identical early B companions displaying broad and shallow lines, and a third O-type star. Because the O star has much sharper line profiles, it is easily disentangled from the close B+B pair using multi-Gaussian profile fitting. Based on the \hea\l4471 to \heb\l4542 line ratio, we assign the O star an O7 spectral sub-type, with the O7.5 type at 1-\s. With no sign of emission lines plus the fact that the observed (diluted) EW of the \heb\l4686 is $\log W_{\lambda4686}=2.64\pm0.06$ (m\AA), we confirm the dwarf luminosity class. The signatures of the fainter companions are not visible in the \heb\l4542 line, but show a faint contribution to \heb\l4686. We thus adopt a B0.5 spectral sub-type for both companions of the O star. Given the object magnitude, we consider that the B+B pair is also formed by two dwarfs. While in reality the two B stars are not exactly identical as indicated by slight differences in their EWs (Fig.~\ref{fig: bd4929}), more data would be needed to refine the classification further. 

The B+B binary has a probable period close to 4~d, but our data do not allow us to put quantitative constraints on the orbit of the tight system. Similarly, the O-type component does not show any significant RV variations and we cannot decide whether the O star is gravitationally linked to the B+B pair or not. We however note that the chance of spurious alignment of an O and a B star in NGC\,6611 is very small. \bd4929 is thus an excellent candidate of hierarchical triple system formed by a close B0.5~V+B0.5~V pair plus a wider O7~V companion.

\begin{figure}
\centering
\includegraphics[width=\columnwidth]{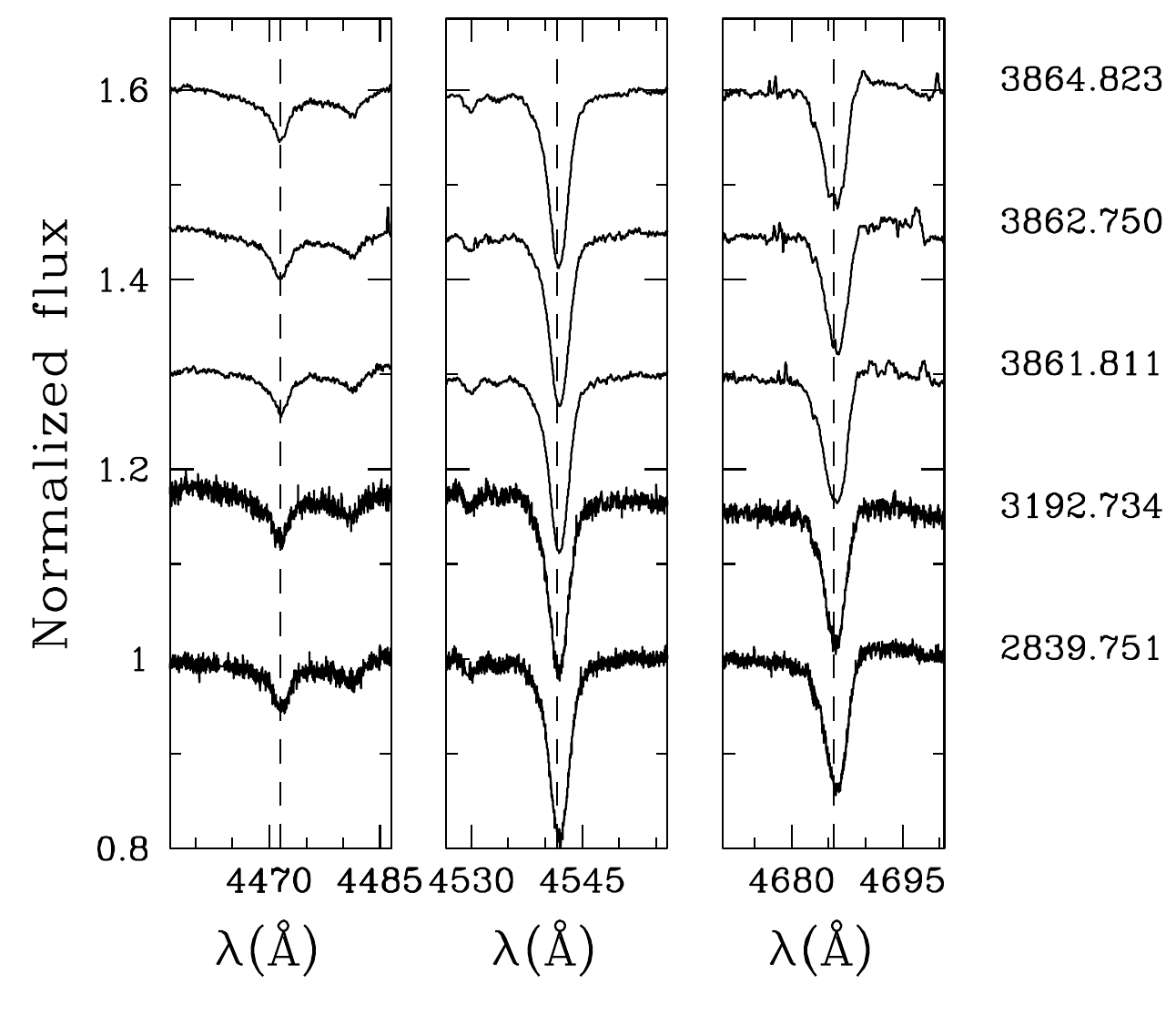}
\caption{{\bf HD\,168076:} \hea\l4471 and \heb\ll4542, 4686 line profiles at different epochs.}
\label{fig: hd076}
\end{figure}

\sss{HD\,168076 (W205)}
Close to the cluster centre, HD\,168076 is the brightest optical source in NGC\,6611. Reported as an SB candidate by \citet{CLL77} based on two discrepant RV points, further investigations by \bmn\ could not confirm its binary nature. We collected only a few spectra in 2006 that show no variability, neither on short time scales nor, by comparison with similar quality data from \esl\ and from May 2009, on longer time scales (Fig.~\ref{fig: hd076}). The EW ratio of the \hea\l4471 and \heb\l4542 lines and the presence of \heb\l4686 strongly in absorption point towards an O4~V((f)) classification, as previously adopted by various authors.

\bmn\ however reported a relatively bright companion at 0.15\arcsec\ from HD\,168076. Given the aperture on the sky of the \feros\ fibre ($\diameter=2.0$\arcsec), the spectrograph thus provides us with a blended spectrum of the two stars and the classification criteria used so far are not applicable. In a second approach, we used the O2-O4 spectroscopic atlas of \citet{WHL02} to refine the spectral classification of HD\,168076. Because of the presence of \ne\ in absorption and of \sid\ in emission with moderate strength, but no \nd\l4058 (Fig.~\ref{fig: hd076spec}), the spectrum of HD\,168076 looks very much like the one of HD\,93128 and we thus adopt an O3.5~V((f+)) spectral classification for the earliest object of the pair.

\citet{DSL06} reported masses of 38 and 16 for both objects, which is likely to have been underestimated given that a typical O3.5~V star is about 52~\msol\ \citep{MSH05}. Assuming that the quoted mass ratio ($q=0.42$) is correct, the secondary component could be an O7.5~V or an O9~III. The expected luminosity ratio would correspond to 5 and 3.5 respectively, which is sufficient in both cases to significantly affect the apparent \hea\ to \heb\ line ratio in the composite spectrum. Because it is unlikely that the secondary evolved more quickly than the primary and because previous mass-transfer episode appears very unlikely in such a young cluster, we finally adopt the O7.5~V classification for the secondary. Under these assumptions, and given the measured angular separation from \bmn, we estimate a minimal orbital period of several hundreds of years.



\subsection{Presumably single stars} \label{ssect: sgl}

\begin{figure}
\centering
\includegraphics[width=\columnwidth]{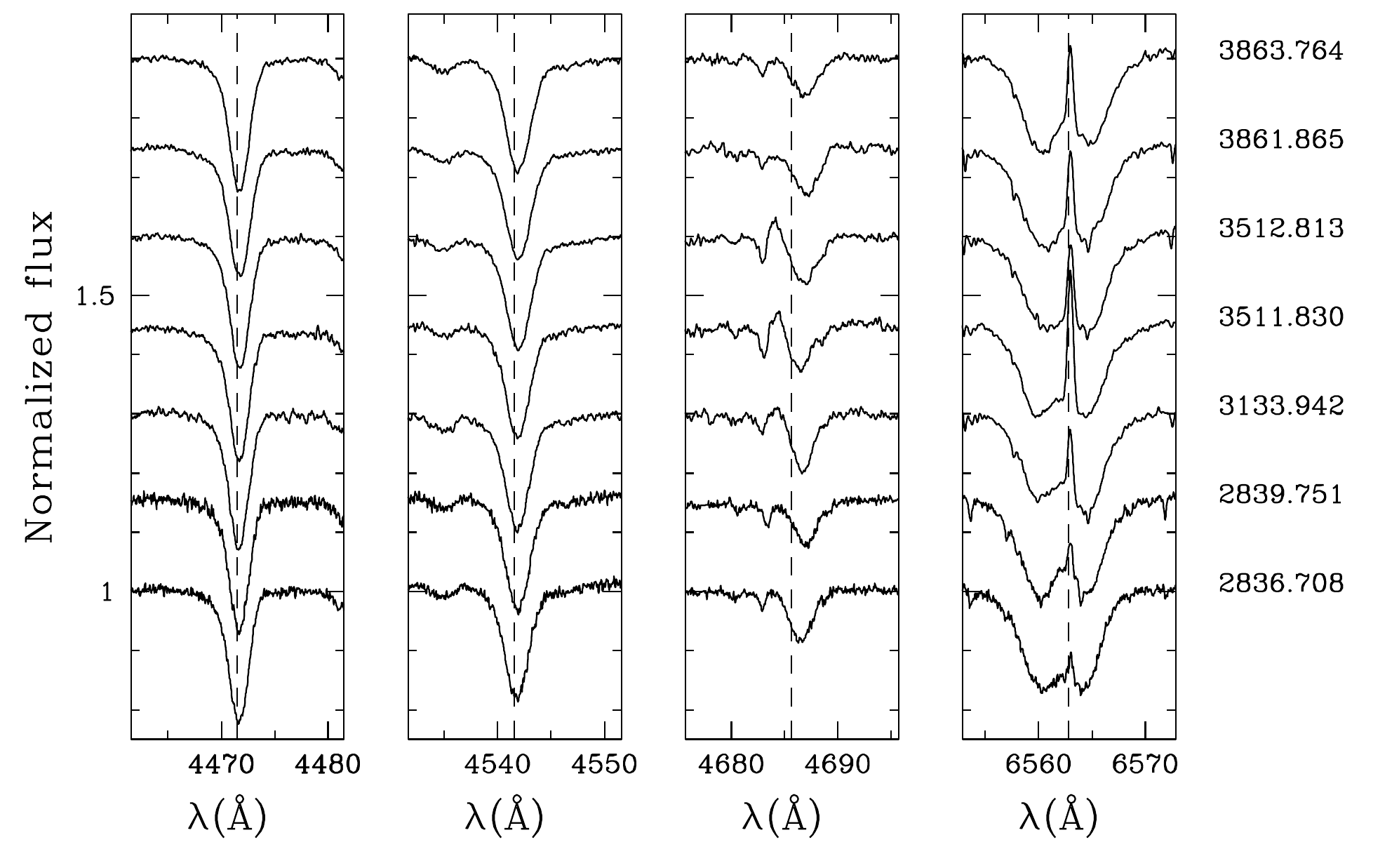}
\caption{{\bf \bd4927:} \hea\l4471, \heb\ll4542, 4686 and \halpha\ line profiles at different epochs. Note the variable emission that partially fills the \heb\l4686 and \halpha\ lines.}
\label{fig: bd4927}
\end{figure}

\sss{\bd4927 (W246)}

Reported as O7~II(f) by \hms, no significant RV variation was detected by \bmn. We collected five \feros\ spectra from 2004 to 2006, plus one spectrum in 2009, and we further included \gir\ data from 2007 in our analysis. With $W_{\lambda 4471} \approx W_{\lambda 4542}$, this object is definitely an O7 star.   Small RV variations ($\sim$10~\kms\ peak-to-peak) are detected, although at the limit of significance, and can easily be mimicked by low-amplitude wind effects. Because $\log W_{\lambda 4686}<2.75$ (m\AA) and because \heb\l4686 and \halpha\ are partly filled with emission (Fig.~\ref{fig: bd4927}), the adopted spectroscopic criteria indicate a giant luminosity class. Yet, with \nc\ll4634-40-42 in emission, the spectrum of \bd4927 is much more alike HD\,151515 (O7~II(f)) than HD\,93522 (O7~III((f))). Combined with the fact that BD$-$13\degr4927 displays a magnitude midway between O7~III and O7~I typical objects \citep{MSH05}, we confirm the O7~II(f) spectral classification. From our campaign, the star is likely to be single.

\begin{figure}
\centering
\includegraphics[width=\columnwidth]{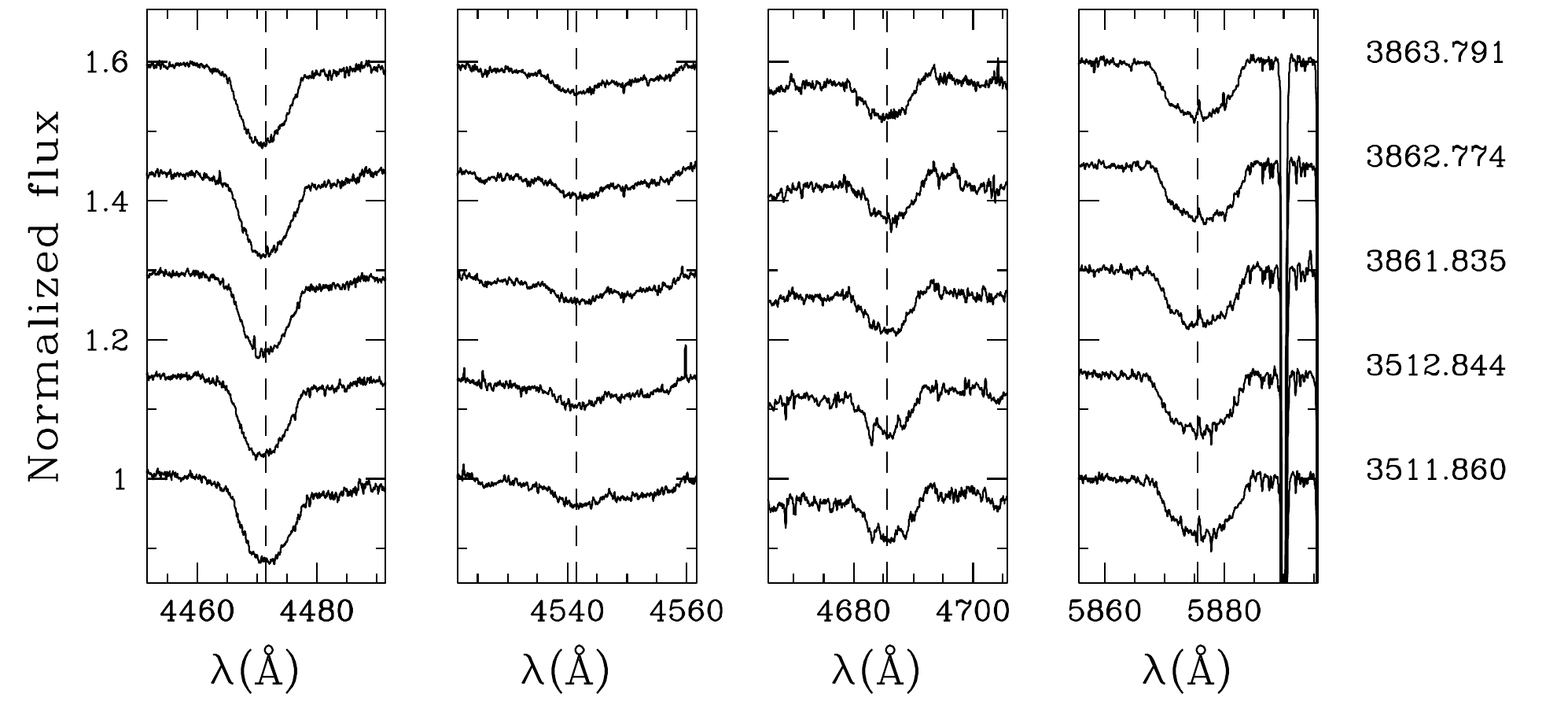}
\caption{{\bf \bd4928:} \hea\ll4471, 5876 and  \heb\ll4542, 4686 line profiles at different epochs. Note the shallow \heb\l4542 line.}
\label{fig: bd4928}
\end{figure}

\sss{\bd4928 (W280)}
Located close to the cluster centre, BD$-$13\degr4928 was reported as a single O9.5~V star by \bmn. We obtained seven additional \feros\ spectra that revealed rotationally broadened lines (Fig.~\ref{fig: bd4928}). Because of this, very few continuum windows were available  in the $\lambda\lambda$4050-4700~\AA\ region of the spectrum. In this range, we used the spectra obtained just before or just after those of \bd4928 to empirically  correct for the instrument response curve. Using this approach, we routinely achieved a precision better than half a percent of the continuum in the response curve correction, only leaving a residual slope in the \bd4928 spectrum. The latter slope, due to the difference of effective temperature between the two objects, was finally removed by fitting a low order polynomial to the continuum of the pre-normalised spectra. 

Minor line profile variations seem to be present but can originate from the object's rapid rotation. Because the lines are rotationally broadened, their profiles deviate significantly from a Gaussian shape. In the present case, we used a rotation profile to measure the Doppler shifts, fitting simultaneously the line centre, its amplitude and the star's projected rotational velocity $v \sin i$. The RVs obtained show a rms dispersion close to 10~\kms\ for rotational velocities in the  range 360-390~\kms. Based on the $W'$ and $W'''$ criteria, we confirm the O9.5~V classification.

\begin{figure}
\centering
\includegraphics[width=\columnwidth]{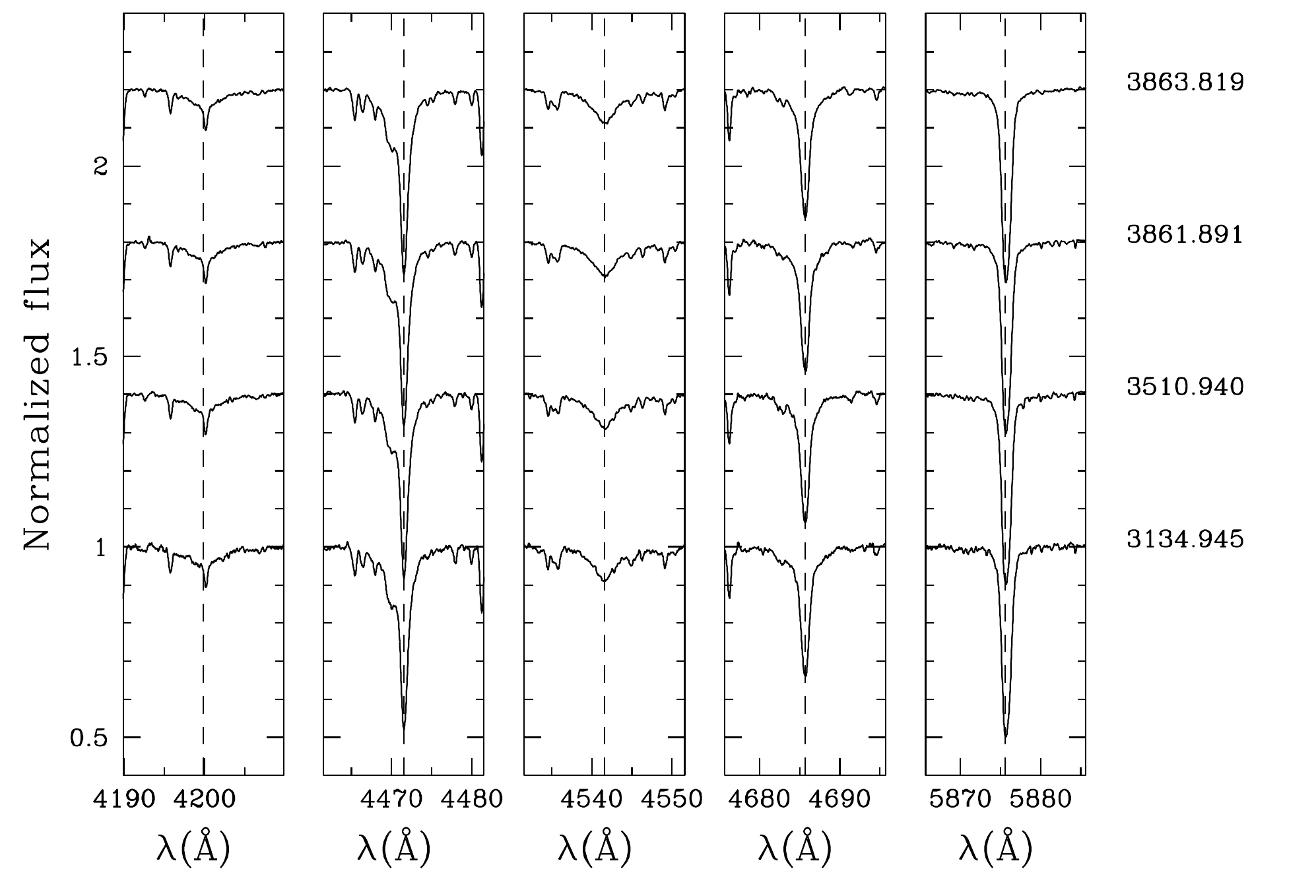}
\caption{{\bf \bd4930:} \hea\ll4471, 5876 and  \heb\ll4200, 4542, 4686 line profiles at different epochs. }
\label{fig: bd4930}
\end{figure}
\begin{figure*}
\centering
\includegraphics[width=15cm]{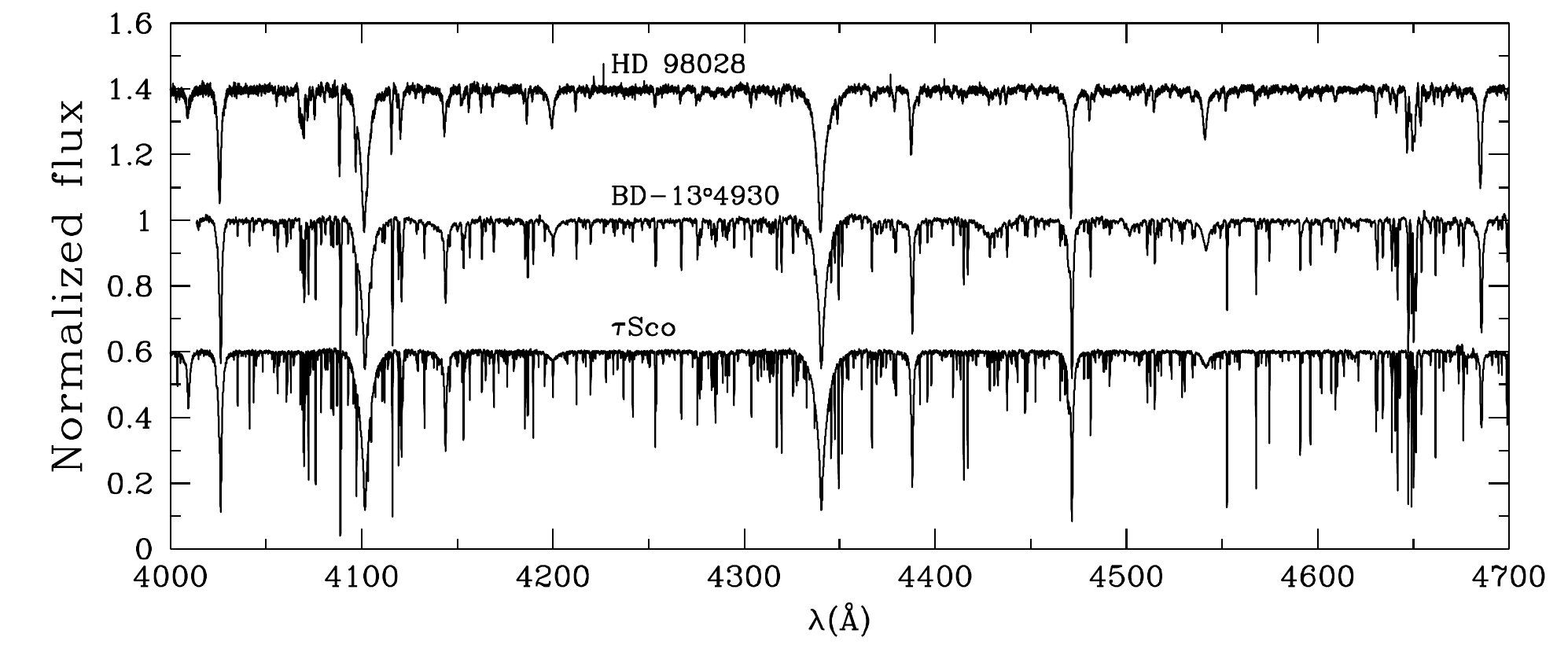}
\caption{{\bf \bd4930} spectrum in the range 4000-4700~\AA\ compared to the spectra of HD\,98028 (O9~V) and \tausco\ (B0.2~V). }
\label{fig: bd4930spec}\end{figure*}

\sss{\bd4930 (W367)}
 So far classified as O9.5~V \phms, \bmn\ noticed that the star had a peculiar velocity compared to the cluster systemic velocity, thus questioning the membership of BD$-$13\degr4930 to \ngc. 

We acquired six \feros\ spectra, complemented by two \gir\ spectra and one \feros\ spectrum from \esl. With an internal RV dispersion of about 1~\kms\ only, we could not detect any variability in our data. \bd4930 is definitely a slow rotator. Most of the lines display a steep Lorentzian profile although some lines, such as \hea\l5876 definitely show a Gaussian profile.  The \heb\ lines further display a broader profile than the \hea\ lines. Using rotationally-broadened profiles for comparison, we estimated that the core of the \hea\ lines is not compatible with a projected rotational velocity larger than 40~\kms. This is in line with earliest measurements by \citet{DSL06} and \citet{HDS07} (see Table~\ref{tab: ID}). The \heb\l4686 line core indicates $v \sin i\approx50$~\kms; the \heb\l5412 line core,  $v \sin i\approx60$~\kms\ while the \heb\l4542 core yields $v \sin i\ga100$~\kms. While we use a purely rotational profile for comparison, we note that the broadening of the \heb\ lines might rather be due to the Stark broadening.

While clearly displaying the \heb\ll4542, 4686 and 5412 lines typical of an O-type star (Fig.~\ref{fig: bd4930}), \bd4930 also displays a large number of metallic sharp lines typical of late-O/early-B stars (Fig.~\ref{fig: bd4930spec}).  Considering only the \hea\ and \heb\ lines, our spectral classification criteria suggest an O8.5~V spectral type, with the O9~V sub-type well within 1-\s. \feros\ spectra of two spectral standards from \citet{WF90}, HD\,93028 (O9~V) and \tausco\ (B0.2~V), are available in the \feros\ archive from ESO. We retrieved both of them and reduced them as described in Sect.~\ref{sect: obs}. Fig.~\ref{fig: bd4930spec} provides a direct comparison of the three spectra. Clearly, \bd4930 displays properties at mid-course between  HD\,93028 and  \tausco. We thus adopt O9.5~V as our final classification.

\section{Observational biases} \label{sect: mc}

\begin{table}
\centering
\begin{minipage}{\columnwidth}
\caption{Binary detection probability for the time sampling associated with different objects (Col.~1) and for various period ranges (Cols. 2 to 5).}
\label{tab: mc_indiv}
\begin{tabular}{ccccc}
\hline
Time sampling & Short   & Intermed. & Long & All \\
          & [2-10d] & [10-365d] & [365-3000d] &[2-3000d] \\   
\hline
\bd4923   & 0.997 &  0.924 &  0.697 &  0.933 \\ 
\bd4927   & 0.994 &  0.897 &  0.695 &  0.925 \\ 
\bd4928   & 0.990 &  0.829 &  0.534 &  0.883 \\ 
\bd4929   & 0.995 &  0.819 &  0.521 &  0.878 \\ 
\bd4930   & 0.993 &  0.911 &  0.690 &  0.928 \\ 
HD\,168075 & 0.995 &  0.913 &  0.667 &  0.848 \\ 
HD\,168076 & 0.995 &  0.925 &  0.690 &  0.932 \\ 
HD\,168137 & 0.995 &  0.942 &  0.727 &  0.943 \\ 
HD\,168183 & 0.991 &  0.876 &  0.508 &  0.891 \\ 
\hline
\end{tabular}
\end{minipage}
\end{table}

\begin{figure}
\centering
\includegraphics[width=\columnwidth]{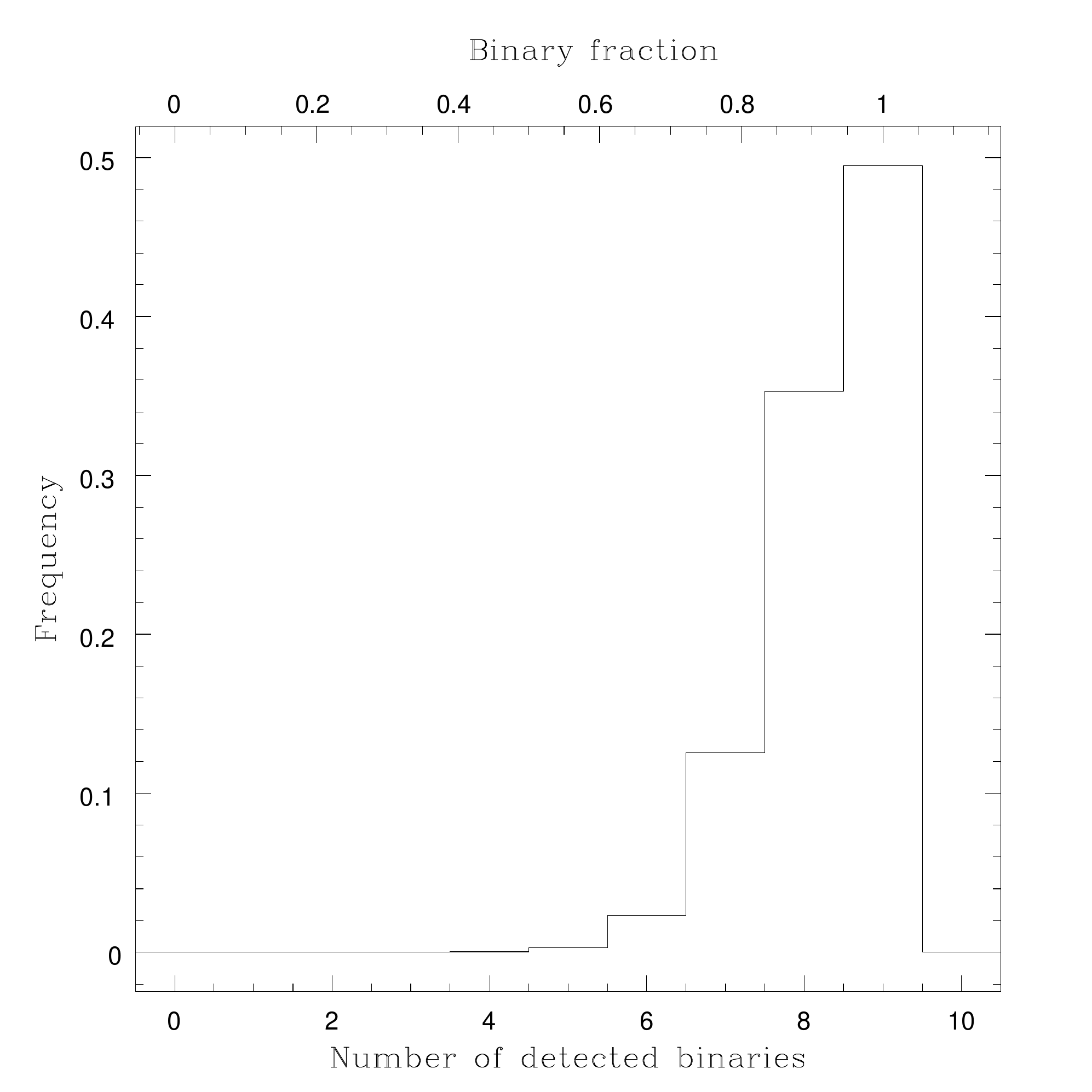}
\caption{Distribution of the number of detected SB systems as obtained from 10\,000 realisations of a cluster with nine O-type binaries.}
\label{fig: mc_all}
\end{figure}

RV techniques alone have proved to be an efficient method to detect high-mass spectroscopic binaries. Yet, the parameter space to probe is very large. It covers 3 to 4 orders of magnitude in period, two orders of magnitude in mass (both for the primary and secondary) and the full range of eccentricities from circular orbits to eccentricities of 0.9 or above. Naturally, RV techniques are not equally sensitive in all the regions of  the parameter space. This not only results from the amplitude of the RV signal but also from the very different time sampling needed to uncover a short-period system with a high mass-ratio  versus a long period, eccentric object with a lower mass secondary component. As a consequence, depending on the probed location in the parameter space, a given time sampling does not allow us to unveil multiple systems with the same probability. 

To better estimate the observational biases of our campaign, we have run a series of Monte-Carlo simulations using two different approaches. In the following simulations, we adopted a  $\Delta$RV threshold for the primary star of 20~\kms\ to consider a binary as detected.


Estimating the detection biases is unfortunately an ill-posed problem as the probability of detection in a given part of the parameter space and for a given observational sampling, needs to be convolved with the distribution of the orbital parameters which, for massive stars, are still poorly constrained. Still, reasonable estimates of the detection biases can be obtained using assumptions that are based on the results of \citetalias{SGN08}. The period distribution is chosen to be bi-uniform in $\log$ scale, between $\log P=0.3$ and 1.0 (days), and between 1.0 and 3.5 (days), with 60\%\ of the systems in the former interval. The eccentricity distribution is taken to be uniform between 0.0 and 0.9, with the additional constraint that the separation at periastron passage should be larger than 20~\rsol\ in order to avoid contact. The mass-ratio distribution is chosen to be uniform between 0.2 and 1.0. 

Adopting the actual observing sampling corresponding to each object, we randomly drew 10\,000 orbital configurations, adopting a random orientation of the orbital plane in space and a random time of periastron passage. Table \ref{tab: mc_indiv} reports the detection probabilities (\pdetect) for short, intermediate and long period binaries, as well as for the full period ranges. It indicates that our campaign is very sensitive in the short-period regime (\pdetect$>0.99$) while its performances remain very good up to one year (\pdetect$\approx0.8-0.9$). Beyond that, the detection probability tends to drop to 0.5-0.6 typically. Yet, given that only 15\%\ of the spectroscopic binary population is expected to have periods between one and ten years, our lower sensitivity in this range has a limited impact on the total detection probability as shown by the last column of Table~\ref{tab: mc_indiv}.

As a second step, to check the consistency of the estimated individual observational biases in the context of a sample of several stars, we randomly drew nine orbital configurations for nine O stars whose masses match the observed primary masses in \ngc. We then applied the existing time sampling of our O-type star observations, randomly associating each time series to an object of the simulated cluster, then deducing an observed binary fraction. Repeating this for 10\,000 simulated clusters, Fig.~\ref{fig: mc_all} shows the distribution of the detected binary fraction. It predicts that our campaign should have been able to achieve a detection rate of $0.93^{+0.07}_{-0.10}$ [0.85 confidence interval], leaving little room for undetected spectroscopic binaries. 

To conclude this section, we note that our simulations do not take into account line blending, so that our adopted detection threshold might be too optimistic for SB2 binaries with similar mass and flux. Yet, a semi-amplitude of the RV-curve of several 10~\kms\ would already result in significant changes ($>20\%$) of the line depths, so that the binary nature of those objects would be unveiled by their line profiles rather that by their RV variations. Only some of the long-period systems ($q>0.8$, $P>1000$~d) would likely remain undetected. This additional shading of the parameter space decreases the total detection probabilities given in Table~\ref{tab: mc_indiv} and Fig.~\ref{fig: mc_all} by a couple of percent, which does not qualitatively affect our conclusions.



\begin{table*}
 \centering
 \begin{minipage}{150mm}
  \caption{Final spectroscopic classification and multiplicity properties of the studied O stars in \ngc. 'SB2O' means that an orbital solution has been obtained; 'SB2E', that the system displays eclipses in addition to its SB2 signature.}
\label{tab: bin}
  \begin{tabular}{@{}llllllllll@{}}
  \hline
Object    & Mult.         & \multicolumn{2}{c}{Sp. Type}             & $q=M_2/M_1$ & $P$            & $e$    \\
          &               & \citeauthor{ESL05mnras} & This work      &             & (d)            &        \\
  \hline            
\multicolumn{7}{c}{Gravitationally-bound systems} \\
  \hline                                          
\bd4923   & SB2O          & O5~V((f+))+O       & O4~V((f+))+O7.5~V   & 0.6        & 13.3           & 0.3    \\
HD\,168075$^a$ & SB2O      & O6-7~V((f))        & O6.5~V((f))+B0-1~V  & 0.487      & 43.6           & 0.17   \\
HD\,168137 & SB2           & O8.5~V             & O7~V+O8~V         & 0.75-0.8     & $>>$30         & undef. \\
HD\,168183 & SB2OE         & B0~III             & O9.5~III+B3-5~V/III & 0.3        & 4.015          & 0.05   \\
  \hline                                                      
\multicolumn{7}{c}{Composite systems} \\                      
  \hline                                                      
\bd4929   & SB3           & O9~V               & O7~V+(B0.5~V+B0.5~V)    & 1.0+(1.0)  & undef.+($\sim$4) & undef. \\
HD\,168076 & composite     & O4III~((f+))       & O3.5~V((f+))+O7.5~V & 0.43           & $>$1.5~10$^5$?   & undef. \\
 \hline
\multicolumn{7}{c}{Single stars}     \\
\hline
\bd4927 & sgl.            & O7~II(f)           & O7~II(f)            & \na& \na & \na  \\
\bd4928 & sgl.            & O9.5~Vn            & O9.5~V              & \na& \na & \na  \\
\bd4930 & sgl.            & O9.7~IIIp          & O9.5~Vp             & \na& \na & \na  \\
\hline
\end{tabular}\\
$a.$ Orbital parameters for HD\,168075 are taken from  \citet{BarbaVina}.
\end{minipage}
\end{table*}

\begin{figure}
\centering
\includegraphics[width=\columnwidth]{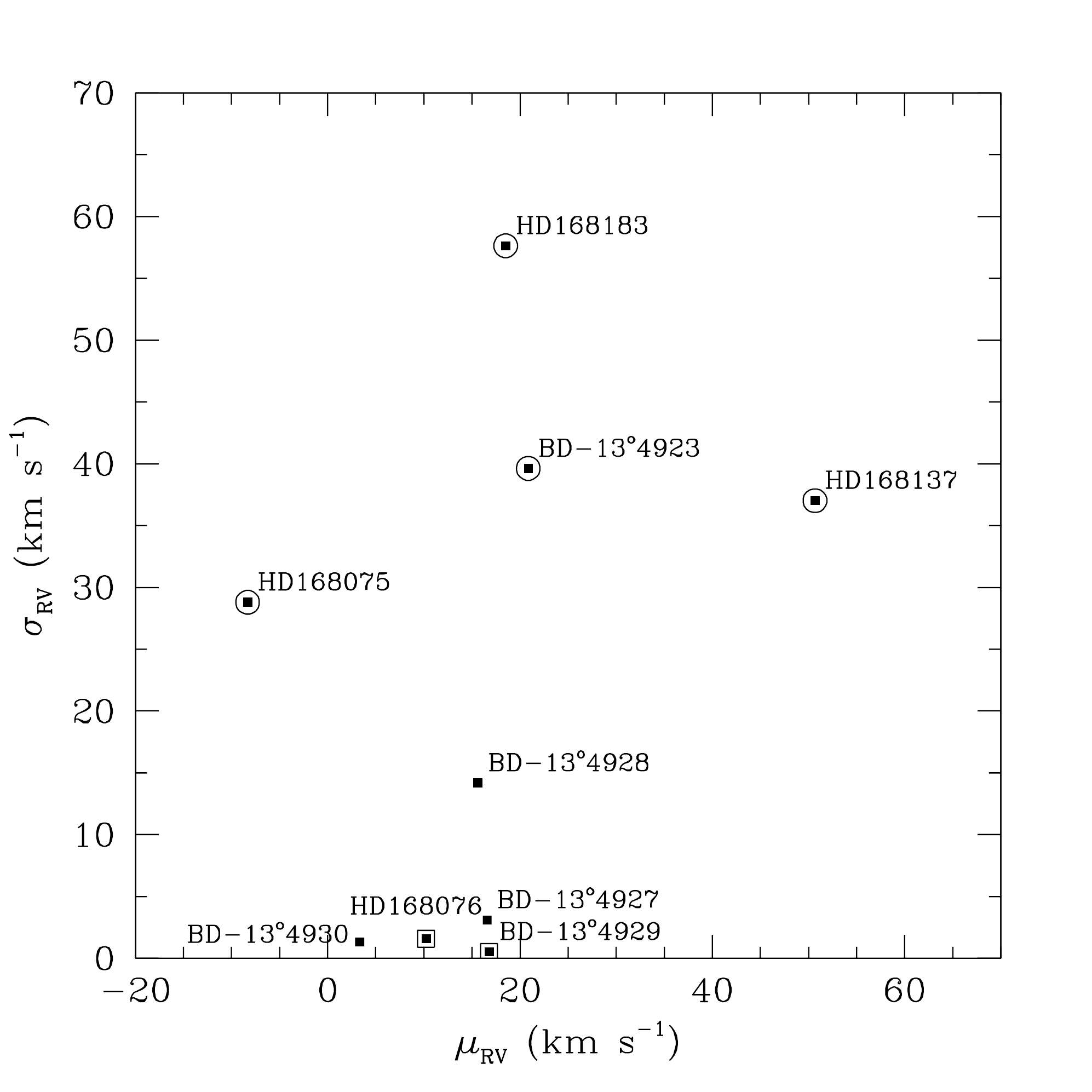}
\caption{Dispersion of the \hea\l5876 RV measurements versus their mean RV. Circles around the $\sigma-\mu$ point indicate the gravitationally bound binaries while squares show the objects with a composite spectrum.  The average systemic velocities have been used instead of the mean RV for the two binaries with orbital solutions : \bd4923 and \hd183.}
\label{fig: mu-sig}
\end{figure}

\section{Discussion} \label{sect: discuss}

\subsection{Membership}
Fig.~\ref{fig: mu-sig} displays the observed 1-\s\ RV dispersions of the \hea\l5876 line in our sample, plotted versus the average RV measurements of the same line. In such a diagram, the vertical axis gives us an indication of the variability while the horizontal axis provides us with some indication of the cluster systemic velocity. Beyond  the four gravitationally bound binaries, it is \bd4928 that shows the largest apparent variability. Yet, as previously discussed, the star is undergoing rapid rotation, thus broadening the spectral lines and making the exact determination of the line centroid less accurate. The remaining stars present an rms dispersion of their RV measurements of a few \kms\ at most and, indeed, all of them are considered to be isolated or very long period objects. 

Both \bd4930 and HD\,168076 show smaller peculiar velocities compared to the average cluster velocity ($\approx17$~\kms). Yet we note that HD\,168076 is a possible long period binary, so that the observed shifts might be additional hints of an ongoing physical process rather than indications of non-membership. Assuming \bd4930 is not a member of \ngc, the following discussion would mostly remain unaffected.

\subsection{Binary fraction}

With four definite SB2 gravitationally bound systems out of nine objects in our sample, the minimal binary fraction is $f_\mathrm{min}\sim0.44$. As discussed in Sect.~\ref{ssect: comp}, two objects display an apparently composite spectrum and could be long period binaries if physically bound. Summing up those binary candidates would increase the binary fraction $f$ up to 0.67.  From Table~\ref{tab: mc_indiv}, one can further conclude that there is little room for non-detection biases for the remaining three stars, making them likely single. For the sake of the present discussion, we will adopt a formal binary fraction of $f=0.55\pm0.11$ in our sample. 

Using the same reasoning as in \citetalias{SGN08}, and assuming that the O stars in our sample are randomly drawn from a binomial parent distribution, we infer a minimal binary fraction for the parent distribution of 0.33 at the 0.01 {\bf significance} level. We note that both the binary fraction and the constraints on the parent distribution parameters are compatible with the results obtained for NGC\,6231 (see  \citetalias{SGN08}). Yet the detailed estimate of the detection biases done in Sect.~\ref{sect: mc} allows us to address another aspect of the parent distribution. Fig.~\ref{fig: mc_all} reveals that the likelihood to detect five or fewer binaries under the hypothesis that all the O stars in NGC\,6611 are binaries amounts to 0.003 only. As a consequence, our observations are not compatible with a contemporaneous multiplicity rate of 100\%.

\subsection{Orbital parameters}

While our campaign cannot allow us to derive the complete set of orbital parameters of all the SB2 systems, the periods and mass-ratios are reasonably well constrained (Table~\ref{tab: bin}). The binary fraction in \ngc\ is similar to the one in NGC\,6231 \citepalias{SGN08}. Yet, \ngc\ lacks the overabundant short-period population seen in NGC\,6231. This however might  result from the small size of our sample. A Kolmogorov-Smirnof test does indeed not authorize to reject the hypothesis that the two samples are taken from the same underlying distribution. Similarly, the distributions of mass ratios in the two clusters are in good agreement and are both compatible with a uniform distribution from $q=1$ to $q=0.2-0.3$ (the observational limit for secondary detection in massive spectroscopic binaries). Finally, we note that the location, in the period-eccentricity diagram, of the systems with known eccentricity would remain well within the area normally populated by O-type binaries. 

\subsection{Companion mass function}

With two O+O and two O+B binaries, plus two additional O+OB candidates, the secondary companion of an O-type star in \ngc\ is strongly biased towards higher masses. Indeed, only three systems, the presumably single stars, could still hide undetected, low-mass companions. As we pointed out in  \citetalias{SGN08}, the mass of an O star companion cannot be randomly drawn from a normal initial mass function. 

\begin{table}
 \centering
 \begin{minipage}{80mm}
  \caption{New spectroscopic distance moduli (DM) for the O stars in this work.}
\label{tab: DM}
  \begin{tabular}{@{}llllllllll@{}}
  \hline
Object    & $E(B-V)$      & $V_0$ & $M_\mathrm{V}^\mathrm{th.}$ &  DM \\
  \hline   
\bd4923   & 1.26$\pm$0.02 & 5.29$\pm$0.15 & -5.85 & 11.14$\pm$0.15 \\
HD\,168076 & 0.71$\pm$0.10 & 5.52$\pm$0.38 & -5.94 & 11.46$\pm$0.38 \\
HD\,168075 & 0.81$\pm$0.11 & 5.74$\pm$0.41 & -5.65 & 11.39$\pm$0.41 \\
HD\,168183 & 0.60$\pm$0.10 & 5.94$\pm$0.38 & -5.20 & 11.14$\pm$0.38 \\
\bd4929   & 0.98$\pm$0.21 & 6.17$\pm$0.81 & -5.24 & 11.41$\pm$0.81 \\
HD\,168137 & 0.68$\pm$0.03 & 6.42$\pm$0.12 & -5.33 & 11.75$\pm$0.12 \\
\bd4927   & 0.94$\pm$0.61 & 7.59$\pm$2.34 & -5.54 & 13.13$\pm$2.34 \\
\bd4928   & 0.76$\pm$0.05 & 7.19$\pm$0.20 & -3.90 & 11.09$\pm$0.20 \\
\bd4930   & 0.60$\pm$0.08 & 7.11$\pm$0.31 & -4.34 & 11.45$\pm$0.31 \\
\hline
\end{tabular}
\end{minipage}
\end{table}

\subsection{Distance}
Using the new spectral classification listed in Table~\ref{tab: bin}, we estimated the spectroscopic distance moduli of each star. This ultimate check allows not only to test the agreement between the spectral types and the object brightnesses, but allows for an independent estimate of the cluster distance. For this purpose, we adopted the calibration of \citet{MaP06} for O stars and the one of \citet{HM84} for B stars. For multiple objects, we weighted the intrinsic color excess $E(B-V)$ and the bolometric correction by the expected contributions of the components to the total flux. Errors were estimated by means of the error propagation, adopting a dispersion corresponding to half a sub-spectral type on the parameters taken from the calibrations. The reddening law is taken to be $A_\mathrm{V} = 3.75 E(B-V)$ \citep{HMS93}. Table~\ref{tab: DM} lists the visual magnitude corrected for reddening ($V_0$), the expected absolute magnitude given the component spectral types ($M_\mathrm{V}^\mathrm{th.}$) and the corresponding distance moduli (DM). Given the estimated error-bars, the latter are in very good agreement with each other. A weighted average
 gives $\overline{DM}=11.4\pm0.1$. This corresponds to a distance to \ngc\ of $d=1.9\pm0.1$~kpc, in excellent agreement with the estimate of \citet{DSL06}.

\section{Summary}\label{sect: ccl}
Using a set of about 100 medium- to high-resolution spectra collected over several years, we revisited the properties of the current O-type star population of \ngc\ in the Eagle Nebula, with a particular emphasis on their spectral classification and multiplicity.
 
We unveiled several new binaries and confirmed two suspected candidates. We proposed the first SB2 solutions for \bd4923 and HD\,168183. We further identified several objects, possible long-period systems, displaying a composite spectrum. We detected the secondary companion signature for all the detected binaries allowing us to put important constraints on the mass-ratio distribution and on the distribution of secondary masses.  As a support to our analysis, we further present a set of Monte-Carlo simulations that assess the binary detection biases of our campaign and show that the latter do not affect our conclusions.

The minimal binary fraction is $f_\mathrm{min}\approx0.44$ but can be as large as 0.67 if the two composite objects are confirmed to be gravitationally bound. In all cases, our current data set excludes a 100\%\ binary fraction in \ngc. Still, up to 75\% of the total O-type star population is to be found in multiple systems. The  period and mass-ratio distributions obtained are both compatible with those derived in \citetalias{SGN08} for NGC\,6231, although the two samples suffer from low-number statistics. All the detected companions are O or early- to mid-B stars, suggesting again that the secondary companion of an O-type star cannot be randomly drawn from a classical mass function but has a mass strongly biased towards the upper range of the mass spectrum. 

Finally, we could not detect any significant difference in the multiplicity properties of the O star population of \ngc\ and NGC\,6231. The environmental properties in which those two populations evolved are however significantly different. On the one hand,  \ngc\ is likely very young, about 1 to 2~Myr old as attested by the presence of two O3-4 stars but no Wolf-Rayet object, and still very dusty. On the other hand, NGC\,6231 is significantly older (around 3-4~Myr) and has already cleared out its dust.

The results presented here and the significant revision of the status of several objects despite the relative proximity of the cluster and the  abundant literature on its O-star population shows the need to pursue our effort on several other nearby clusters. Only upon completion of our campaign will we be able to draw a global picture of the properties of the O-type star population in the nearby Universe.


\section*{Acknowledgments}
The authors are grateful to R. Barb\'a for sharing results before publication and to the referee, D. Gies, for a careful revision of the manuscript. This paper relies on data taken at the La Silla-Paranal Observatory under program IDs  71.C-0513, 171.D-0237, 073.D-0234, 073.D-0609, 075.D-0061, 075.D-0369, 077.D-0146, 079.D-0564 and 082.D-0136(A). This paper also made use of the ADS, of the SIMBAD and WEBDA databases and of the Vizier catalogue access tool (CDS, Strasbourg, France).


\bibliographystyle{mn2e}
\bibliography{/home/hsana/LITERATURE/literature}

\bsp

\label{lastpage}

\end{document}